\documentclass[
twoside,
10pt,
twocolumn,
]{article}
\usepackage[utf8x]{inputenc}
\usepackage{amsmath,amsfonts,amsthm,amssymb}
\usepackage[table,dvipsnames]{xcolor}
\usepackage[final]{graphicx}
\usepackage{subfigure}

\usepackage{cases}
\usepackage{multirow}
\usepackage{url}
\usepackage{hyperref}
\usepackage{booktabs}
\usepackage{array}
\usepackage{tikz}
\usetikzlibrary{shapes.geometric,positioning}
\usepackage{multirow}
\newcolumntype{C}[1]{>{\centering\arraybackslash}m{#1}}

\usepackage{dashrule}
\usepackage[
 capitalise
]{cleveref}
\usepackage{bm}
\usepackage{abstract}

\usepackage{lmodern}
\newtheorem{defn}{Definition}[section]

\usepackage{xcolor}
\definecolor{myGreen}{rgb}{0,0.8,0}
\definecolor{myMagenta}{rgb}{1,0,1}

\newcommand{\EE}[1]{\mathbb{E}\left[#1\right]}
\newcommand{\erf}{\text{erf}}

\hypersetup{
	backref=true,            %permet d'ajouter des liens dans...
	pagebackref=true,    %...les bibliographies
	hyperindex=true,      %ajoute des liens dans les index.
	colorlinks=true,        %colorise les liens
	breaklinks=true,       %permet le retour àla ligne dans les liens trop longs
	urlcolor= blue,        %couleur des hyperliens
	linkcolor= blue,        %couleur des liens internes
	citecolor=Green,
	bookmarks=true,       %cree des signets pour Acrobat
	bookmarksopen=false  %si les signets Acrobat sont crees,
}

\usepackage{geometry}

\graphicspath{{Fig_Article_NES_InfluEps/}}



\usepackage[
font={small},
labelfont=bf,
labelsep=period]{caption}

\usepackage{sectsty}
\sectionfont{
\normalsize
}

\subsectionfont{
\normalsize\normalfont\itshape
}

\subsubsectionfont{
\normalfont
\itshape
}

\usepackage{fancyhdr}
\fancyheadoffset[LE,RO]{0cm}

\usepackage{titling}
\usepackage[auth-lg]{authblk}
\title{Analytical prediction of delayed Hopf bifurcations in a simplified stochastic model of reed musical instruments}

\author{Baptiste Bergeot$^1${\thanks{Corresponding author: \texttt{baptiste.bergeot@insa-cvl.fr}}}
\, and Christophe Vergez$^2$}
\affil{$^1$INSA CVL, Univ. Orl\'{e}ans, Univ. Tours, LaM\'{e} EA 7494, F-41034, 3 Rue de la Chocolaterie, CS 23410, 41034 Blois Cedex, France\\
$^2$ Aix Marseille Univ., CNRS, Centrale Marseille, LMA UMR 7031, Marseille, France}

\date{13 January 2022}
\begin{document}

\twocolumn[
\maketitle

\begin{onecolabstract}
This paper investigates the dynamic behavior of a simplified single reed instrument model subject to a stochastic forcing of white noise type when one of its bifurcation parameters (the dimensionless blowing pressure) increases linearly over time and crosses the Hopf bifurcation point of its trivial equilibrium position. The stochastic slow dynamics of the model is first obtained by means of the stochastic averaging method. The resulting averaged system reduces to a non-autonomous one-dimensional Itô stochastic differential equation governing the time evolution of the mouthpiece pressure amplitude. Under relevant approximations the latter is solved analytically treating separately cases where noise can be ignored and cases where it cannot. From that, two analytical expressions of the bifurcation parameter value for which the mouthpiece pressure amplitude gets its initial value back are deduced. These special values of the bifurcation parameter characterize the effective appearance of sound in the instrument and are called deterministic dynamic bifurcation point if the noise can be neglected and stochastic dynamic bifurcation point otherwise. Finally, for illustration and validation purposes, the analytical results are compared with direct numerical integration of the model in both deterministic and stochastic situations. In each considered case, a good agreement is observed between theoretical results and numerical simulations, which validates the proposed analysis.

\vspace{0.2cm}


\vspace{0.5cm}

\end{onecolabstract}
]
\saythanks

\section*{List of main symbols}

The next list is not exhaustive, it describes only the main symbols that will be later used within the body of the document.

\begin{description}

\item[$p_n$, $p$, $u$]
Physical unknowns in the reed instrument model ($n^\text{th}$ modal component, pressure and volume flow at the input of the instrument respectively).

\item[$x_t, x_y$]
Amplitude of $p$ in the amplitude/phase representation (different subscript is used to stress the dependance).

\item[$A(y)$]
Intermediate quantity used throughout the paper and defined by Eq.~\eqref{eq:defA}..

\item[$\gamma$, $\gamma_t$]
Dimensionless blowing pressure (the subscript $()_t$ indicates that $\gamma$ varies with time).

\item[$\hat{\gamma}^{st}$]
Value of $\gamma$ corresponding to a static Hopf bifurcation.

\item[$y_t$]
Time-varying dimensionless blowing overpressure with respect to the static Hopf bifurcation ($= \gamma_t - \hat{\gamma}^{st}$).

\item[$\hat{y}_\text{det}^\text{dyn}$]
Analytical prediction of the value of $y$ at which the system undergoes a dynamic Hopf bifurcation in the deterministic case.

\item[$y_0$]
Value of $y_t$ at $t=0$.

\item[$\hat{y}_\text{stoch,a}^\text{dyn}$]
Analytical prediction of the value of $y$ at which the system undergoes a dynamic Hopf bifurcation in the stochastic case (the subscript $()_\text{a}$ means first level of approximation).

\item[$\hat{y}_\text{stoch,b}^\text{dyn}$]
Analytical prediction of the value of $y$ at which the system undergoes a dynamic Hopf bifurcation in the stochastic case (the subscript $()_\text{b}$ means second level of approximation, less accurate than $\hat{y}_\text{stoch,a}^\text{dyn}$).

\item[$\nu, \sigma$]
Magnitude of the stochastic forcing ($\sigma = \nu / \sqrt{2}$).

\item[$\epsilon$]
Rate of linear increase of the bifurcation parameter.

\item[$\alpha_i, \omega_i, F_i$]
Modal parameters of the $i^{th}$ mode (damping, eigenfrequency, modal factor respectively).

\item[$\zeta$]
Embouchure parameter (physical model of clarinet).

\end{description}

%-------------------------------------------------------------------------------------------------%
%-------------------------------------------------------------------------------------------------%
% Inroduction
%-------------------------------------------------------------------------------------------------%
%-------------------------------------------------------------------------------------------------%
\section{Introduction}\label{sec:intro}

Musical instruments are nonlinear dynamical systems. An important specificity is that sound production in a musical context corresponds to a time-varying control. Indeed, control parameters are modified continuously by the instrument player. To give just one example, wind instruments players control air pressure in their mouth with variations over time finely tuned to produce the desired sound effect. On the other hand, when studying the corresponding mathematical models of sound production by musical instruments, the control parameters are considered constant over time most of the time. In this respect, the instruments are modeled by autonomous nonlinear systems of differential equations (ODEs) having, among other solutions, at least one trivial equilibrium position (corresponding to silence) and periodic solutions (corresponding to musical notes). The mouth pressure is then a bifurcation parameter of this ODEs system: when it is small the trivial solution is stable and, beyond a precise value called static bifurcation point (“static” because the bifurcation parameter is constant over time), the trivial solution loses its stability through a Hopf bifurcation giving rise to a periodic solution. In musical acoustics literature, the static bifurcation point of the trivial solution is often called oscillation threshold~\cite{dalmont:3294}. In this “static” context, periodic solutions and their stability can be also determined using for example the harmonic balance method~\cite{Freour2020,Colinot2020} or orthogonal collocation~\cite{Terrien2013,Gilbert2020}. Control parameters are changed over time only in the very special context of time simulations, e.g. when the objective is sound synthesis, not model behavior analysis.

The purpose of this paper is not to question these methods (with constant control parameters) which have shown their relevance (experimentally and numerically) when considering steady-state oscillation regimes (i.e. excluding all forms of transients). On the contrary, it is more a question of shedding light on the appearance of a sound when a time-varying control parameter (more specifically, the mouth pressure) crosses the static (Hopf) bifurcation point, as it is the case in playing situations. Moreover, experiments with a blowing machine and a clarinet-like instrument have shown in this case the existence of a delay at the bifurcation~\cite{Jasa2013BBergeot}: when the pressure in the musician's mouth is increased linearly, the start of oscillations is observed for a pressure value greater than the static bifurcation point. This particular value of the mouth pressure for which the oscillations actually appear is called dynamic bifurcation point.

An analytical study to explain this observation has already been carried out on a discrete-time model of the clarinet\footnote{In this case the instrument is modeled by a difference equation similar to the logistic map. This model has been thoroughly studied in musical acoustics for control parameters constant over time~\cite{dalmont:1173,NonLin_Tail_2010,Taillard2015}.}~\cite{BergeotNLD2012,BergeotNLD2012b}. Considering a blowing pressure linearly increasing over time, the following results have been obtained in these works, which are typical of what is known in the field of dynamical bifurcation of discrete-time systems. In fact, the behavior of the bifurcation delay depends on the possibility of ignoring or not the presence of noise in the model. We are referring here only to the presence of an additive white noise. Sometimes even a noise with a very low amplitude coming from the rounding errors of the computer in numerical simulations must be taken into account. If the noise can be ignored the bifurcation delay depends only on the initial value of the linearly increasing mouth pressure, the farthest it starts below the static bifurcation point, the larger the delay. If the noise is no longer negligible, then the bifurcation delay loses the dependence on the initial condition and logically becomes dependent on the noise level but also on the increase rate of the mouth pressure. In this case, the bifurcation delay decreases with the noise level and increases with the increase rate of the mouth pressure. These properties of the bifurcation delay shown in these works are also found in other works in applied mathematics and physics whether for discrete or continuous-time systems with an additive white noise~\cite{Baesens1991,Baesens1991Noise,Jansons1998aa,BGbook}. The strong dependence of the bifurcation delay on noise makes essential to take it into account in the modeling a real life system like a musical instrument.

The discrete-time clarinet model considered in our previous cited works is known for its simplicity and its ability to explain some phenomena observed experimentally. However, the simple formulation of this model is achieved under very strict simplifying hypotheses. The most important ones are frequency independent losses, a cylindrical geometry, idealized radiation ... 

Another approach, based on continuous-time models including ODE’s, allows an easier consideration of refinements in the equations, leading to models with different levels of complexity~\cite{Silva2014}. The simplest model consists in retaining only one acoustic mode of the air-column in the instrument while assuming that the cane reed is driven instantaneously by the pressure difference between the mouth of the player and the input of the instrument~\cite{ChaiKergoEn2016}. The work presented in this paper is based on this model which is here considered with a blowing pressure linearly increasing over time and a stochastic forcing of white noise type.

The questions addressed in this paper are: How does the continuous-time clarinet model behave when the blowing pressure increases linearly, compared to the classical case of a blowing pressure constant over time? How do the results change when random fluctuations in the source term of the model are added (which may represent an idealized version of the turbulent noise due to the airflow)? More precisely, through relevant approximations, the objective is to solve the studied model analytically by considering the general framework of the stochastic differential equations in order to predict the dynamic bifurcation points of the model.

The paper is organized as follows. The single reed instrument model is presented in Section~\ref{sec:model}. Section~\ref{sec:modelA} gives the equations of motion of the classical deterministic one-mode model of single reed instruments and recalls the expression of its static bifurcation point. Section~\ref{sec:stocmod} presents the model with, in addition, a stochastic forcing of white noise type, a linearly increasing blowing pressure and pertinent rescaling. In Section~\ref{sec:SSF}, the stochastic averaging method is used to derive the slow dynamics of the model. Through simplifying assumptions, the slow dynamics is solved in Section~\ref{sec:Analy}. The method is based on treating separately cases where noise can be ignored and cases where it cannot. An analytical expression of the dynamic bifurcation point is obtained in each case. In Section~\ref{Numerical_results}, for illustration and validation purposes, these analytical results are compared with direct numerical integration of the model. Finally, concluding remarks and some perspectives are given in Section~\ref{sec:CCL}.

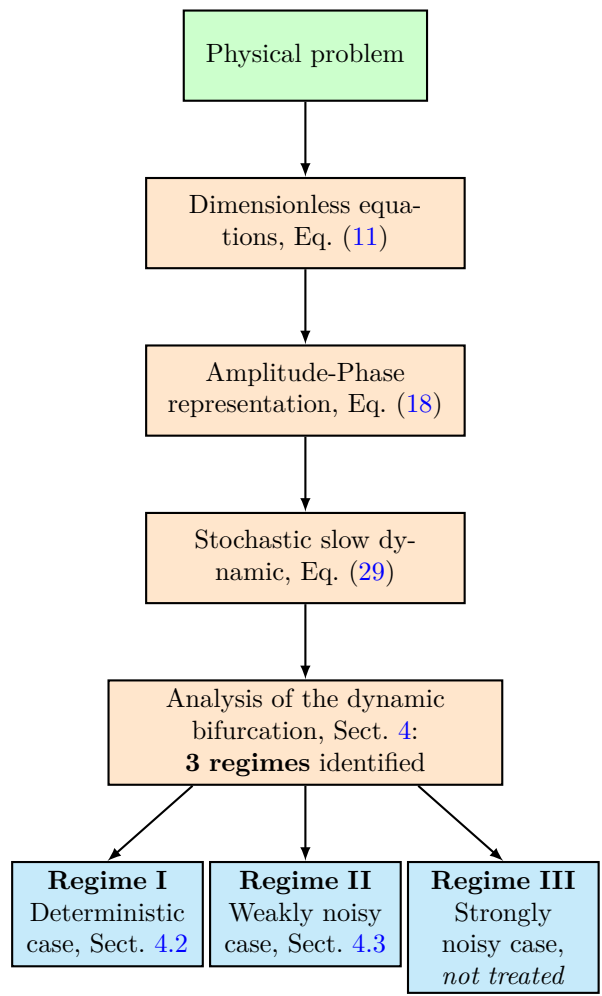
\begin{figure}[t!]
\centering
\begin{tikzpicture}[thick,
	every text node part/.style={align=center},
	gain/.style = {
		draw, 
		isosceles triangle,
		isosceles triangle apex angle=60,
		minimum height = 1.5em,
		outer sep=0},
	TFblock/.style= {
		draw,  
		minimum size=1.2cm}]

\node [TFblock,fill=green!20,text width=3cm] (b) {Physical problem};

% Phase Compensation Block
\node [TFblock,fill=orange!20,text width=4cm] (c) [below =of b] {Dimensionless equations, Eq.~\eqref{eq:modclarstoch2}};
\node [TFblock,fill=orange!20,text width=4cm] (d) [below  =of c] {Amplitude-Phase representation, Eq.~\eqref{eq:ampphaseCar1}
};
\node [TFblock,fill=orange!20,text width=4cm] (e) [below  =of d] {Stochastic slow dynamic, Eq.~\eqref{eq:VdP1Moy3Stochay}
};
\node [TFblock,fill=orange!20,text width=5cm] (f) [below  =of e] {Analysis of the dynamic
%Pitchfork
bifurcation, Sect.~\ref{sec:Analy}: \textbf{3~regimes} identified
};
\node [TFblock,fill=cyan!20,text width=2.3cm] (g) [below  =of f] {\textbf{Regime II} \\ Weakly noisy case, Sect.~\ref{sec:AnaStoch}
};
\node [TFblock,fill=cyan!20,text width=2.3cm] (h) [below  =of f.south west] {\textbf{Regime I} \\ Deterministic case, Sect.~\ref{sec:AnaDet}
};
\node [TFblock,fill=cyan!20,text width=2.3cm] (i) [below  =of f.south east] {\textbf{Regime III}  \\ Strongly noisy case,  \textit{not treated}};

% Arrows
\draw[-latex] 
	(b.south) edge (c.north)
	(c.south) edge (d.north)
	(d.south) edge (e.north)
	(e.south) edge (f.north)
	(f.south) edge (g.north)
	(f.south)++(-1.5,0) edge (h.north)
	(f.south)++(1.5,0)-- (i.north);
\end{tikzpicture}
\caption{Main steps of the proposed approach as a block diagram.}
\label{fig:steps}
\end{figure}

The main steps of the proposed approach are summarized in Fig.~\ref{fig:steps}.

%-------------------------------------------------------------------------------------------------%
%-------------------------------------------------------------------------------------------------%
% Section
%-------------------------------------------------------------------------------------------------%
%-------------------------------------------------------------------------------------------------%
\section{Single reed instrument model with a white noise forcing}
\label{sec:model}

%_____________________________Subsection_____________________________%
\subsection{Deterministic model of single reed instrument and its static bifurcation point}
\label{sec:modelA}

In this section the single reed instrument model is recalled as it is generally discussed in the literature, i.e. it is deterministic and the control parameters are constant over time.

Sound production by single reed instruments is classically modeled through the nonlinear coupling of two linear sub-systems~\cite{Benade_Book,ThePhyOfMusInst,ChaiKergoEn2016}: the cane reed and the air-column inside the instrument. While blowing air through the reed channel into the instrument, the musician provides a quasi-static source of energy. The instrument and the player constitute an autonomous dynamical system.  When the trivial equilibrium solution of this system becomes unstable, a sound is produced~\cite{WilJASA1974,Fletcher199085,SilvaJASA2008}.  

Since the lowest resonance frequency of the reed is one order of magnitude higher than the sound frequency for many notes, the reed is often modeled as a lossless stiffness spring~\cite{backus:305,OllivActAc2004}. Therefore, the position of the reed relative to rest (which determines the opening of the reed channel) is proportional to the pressure drop across the reed, i.e. the pressure difference between the mouth and the mouthpiece of the instrument. The linear pressure response of the air column $P$ to the volume flow $U$ through the reed channel is given in the frequency domain through the input impedance of the air column~$Z$:
\begin{equation}
P(\omega) = Z(\omega) U(\omega),
\end{equation}
where $\omega$ is the angular frequency. The contribution at the input of the instrument of the (infinite) series of modes of the air column is taken into account in $Z(\omega)$. For computational reasons, the series is truncated to $N$ modes, where $N$ is an integer:
\begin{equation}\label{e:Zfreq}
Z(\omega) = \sum_{n=1}^{N} F_n \frac{j\omega}{\omega_n^2 + j \omega\alpha_n\omega_n-\omega^2},
\end{equation}
with $F_n$, $\omega_n$ and $\alpha_n$  the modal parameters, respectively the modal factor, the resonance angular frequency and the inverse of the quality factor of the $n^\text{th}$ peak of the impedance (corresponding to the $n^\text{th}$ mode of the air column). Eq.~\eqref{e:Zfreq} can be written in the time domain:
\begin{equation}
\ddot{p}_n+ \alpha_n \omega_n \dot{p}_n + \omega_n^2 p_n=F_n\dot{u}, \qquad \forall n\in [1,N],
\label{e:pn}
\end{equation}
with $u$ the inverse Fourier transform of $U$ and $p_n$ is such that $p = \sum_{i=1}^{N} p_n$, where $p$ is the inverse Fourier transform of $P~$\cite{SilvaJASA2008} and corresponds to the time evolution of the mouthpiece pressure.

The volume flow through the reed channel is related nonlinearly to the reed channel opening and the pressure difference between the mouth and the mouthpiece~\cite{MOMIchap7,dalmont:2253}. A polynomial expansion of this relation is often written in the neighborhood of the equilibrium solution (i.e. the mean flow)~\cite{MechOfMusInst}:
\begin{equation}
u(t) = u_{eq} + c_1 p(t) + c_2 p(t)^2 + c_3 p(t)^3,
\label{e:debit}
\end{equation}
with $u_{eq}=\zeta(1-\gamma)\sqrt{\gamma}$ the mean volume flow, $c_1=\zeta \frac{3\gamma-1}{2\gamma^{\frac{1}{2}}}$, $c_2=-\zeta \frac{3\gamma+1}{8\gamma^{\frac{3}{2}}}$ and $c_3=-\zeta \frac{\gamma+1}{16\gamma^{\frac{5}{2}}}$, where $\gamma$ is the dimensionless pressure in the mouth of the musician and $\zeta$ a dimensionless parameter accounting for many embouchure parameters. By Eq.~\eqref{e:debit}, Eq.~\eqref{e:pn} can be written using only the pressure $p$ as follows
\begin{equation}
\ddot{p}_n+ \alpha_n \omega_n \dot{p}_n + \omega_n^2 p_n+F_n \dot{p} f(p,\gamma)=0, \quad \forall n\in [1,N],
\label{e:pn2}
\end{equation}
where $f(p,\gamma)=-\frac{\partial u}{\partial p}$.

A minimal model of a reed instrument including a single mode of the air-column is obtained by stating $N=1$. In this case \eqref{e:pn2} becomes
\begin{equation}
\ddot{p}+ \alpha_1 \omega_1 \dot{p} + \omega_1^2 p+F_1\dot{p}f(p,\gamma)=0
\label{eq:VdP1Mode1}
\end{equation}
In this case ($N=1$), since $p_1 = p$, $p_1$ is replaced by $p$ in Eq.~\eqref{eq:VdP1Mode1}.
Note that $p$ and $u$ are dimensionless and $F_1$ unit is s$^{-1}$.  This is clearly a minimal yet useful model of sound production in reed instruments. Indeed, it takes into account the two main control parameters adjusted by the musician and describes the physical mechanism through which sound emerges from equilibrium (i.e. silence) when a resonance of the air column is excited by an incoming flow.

In this paper the bifurcation parameter under consideration is $\gamma$. The stability of the trivial equilibrium solution $p=0$ with respect to $\gamma$ is classically analyzed by looking at the sign of the eigenvalues real parts of the Jacobian matrix of Eq.~\eqref{eq:VdP1Mode1} written in the state-space form. This leads to the following expression for the static Hopf bifurcation point
\begin{equation}
\hat{\gamma}^{\text{st}}=
\frac{1}{3}+
\frac{2 \alpha _1 \omega _1 \left(\alpha _1 \omega _1+\sqrt{\alpha _1^2 \omega _1^2+3 \zeta ^2 F_1^2}\right)}{9 \zeta ^2 F_1^2}
\label{eq:statbifpt}
\end{equation}
corresponding to the value of $\gamma$ for which the two complex conjugate eigenvalues become with positive real parts. Note that in the lossless case (i.e. $\alpha_1=0$) the static bifurcation parameter is $\hat{\gamma}^{\text{st}}=
\frac{1}{3}$. Static bifurcation point is known in the literature of acoustics of musical instruments as the "oscillation threshold" of the instrument. Expressions equivalent to Eq.~\eqref{eq:statbifpt} are given for example by Kergomard \textit{et al.}~\cite{Kergomard2000} and Silva \textit{et al.}~\cite{Silva2008}. 

Compared to the model given by \eqref{eq:VdP1Mode1}, this article studies the case of a linearly increasing blowing pressure $\gamma$, with additional stochastic excitation as detailed in Sect.~\ref{sec:stocmod}.

%_____________________________Subsection_____________________________%
\subsection{One-mode stochastic single-reed instrument model with a linearly increasing blowing pressure}
\label{sec:stocmod}

Hereafter, the formalism of stochastic differential equations is used in the framework of the Itô stochastic calculus. A good description of these concepts can found for example in~\cite{bernt2003}.

The one-mode model described by Eq.~\eqref{eq:VdP1Mode1} is now considered with a linearly increasing blowing pressure $\gamma=\hat{\epsilon} t+\gamma_0$. The variation of the blowing pressure is assumed to be small during a period $T_1=\frac{2 \pi}{\omega_1}$ which is ensured by $0<\hat{\epsilon}\ll 1$. Because one assumes also that $\omega_1\gg1$ (without loss a generality in musical context) the time variation of the mouth pressure $\gamma$ is neglected in the time derivative of the mean flow~\eqref{e:debit}. Moreover, the model is now forced by a stochastic excitation $\hat{\nu}\xi_t$ where $\hat{\nu}$ is the noise level and $\xi_t$ is assumed to be a unitary idealized white noise process with zero mean, i.e.
\begin{equation}
\EE{\xi_t}=0 \quad \text{and} \quad \EE{\xi_t\xi_{t+\tau}}=\delta(\tau)
\label{eq:white1}
\end{equation}
where $\delta$ is the Dirac delta function and $\EE{\{.\}}$ denotes the ensemble average\footnote{We recall that the ensemble average consists in repeating the same measurement many times, and in calculating the average over them.} of $\{.\}$. Finally, the one-mode stochastic model is written as follows
\begin{equation}
\ddot{p}_t+ \alpha_1 \omega_1 \dot{p}_t + \omega_1^2 p_t+F_1\dot{p}_t f(p_t,\gamma_t)
=\hat{\nu}\xi_t
\label{eq:VdP1Stoch1}
\end{equation}
where the subscript $t$ is used to show the stochastic nature of the differential equation. This is the classical notation in the framework of stochastic differential equations. 

From the physical point of view, Eq.~\eqref{eq:VdP1Stoch1} can be interpreted as a simple model of sound production in reed instruments with turbulent noise contribution taken into account as an additive stochastic noise in the source term. 

To facilitate the mathematical developments of the following sections, transformations are performed within Eq.~\eqref{eq:VdP1Stoch1}. First, a new bifurcation parameter $y_t=\gamma_t-\hat{\gamma}^{\text{st}}$ is considered to obtain a system whose trivial solution becomes unstable at $y=0$ (i.e. $\hat{y}^{\text{st}}=0$). Then the time rescaling $t\rightarrow t'=\omega_1 t$ is introduced. Recalling that a normalized white noise $\xi_t$ is defined as the time derivative of the normalized Wiener process $W_t$ and using the \textit{scaling property of Wiener process} (see \cite{bernt2003}, Chap.~2) which states that $W_{\frac{t}{\omega_1}}$ and $\frac{1}{\sqrt{\omega_1}}W_{t}$ are equivalent, we have
\begin{equation}
\xi_{t}=\frac{dW_{t}}{d t}
\sim\omega_1\frac{dW_{{\frac{t'}{\omega_1}}}}{d t'}
\sim\sqrt{\omega_1}\frac{dW_{t'}}{d t'}
=\sqrt{\omega_1}\xi_{t'}.
\end{equation}
Therefore, by using also $\dot{\{\}}$ for the derivation with respect to $t'$ and denoting $t'$ by $t$ for the sake of conciseness, Eq.~\eqref{eq:VdP1Stoch1} takes the form of the following stochastically forced self-excited oscillator
\begin{equation}
\ddot{p}_t
+h\left(p_t,\dot{p}_t,y_t\right)
+p_t
=\nu\xi_t
\label{eq:modclarstoch2}
\end{equation}
where
\begin{equation}
h\left(p_t,\dot{p}_t,y_t\right)=\alpha_1 \dot{p}_t+\frac{F_1}{\omega_1}\dot{p}_tf\left(p_t,y_t+\hat{\gamma}^{\text{st}}\right)
\label{eq:exph1}
\end{equation}
and
$
\nu=\frac{\hat{\nu}}{\omega_1^{3/2}}.
$
The time evolution of $y_t$ is given by
\begin{equation}
y_t=\epsilon t+y_0
\end{equation}
where
$y_0=\gamma_0-\hat{\gamma}^{\text{st}}$ and $\epsilon=\frac{\hat{\epsilon}}{\omega_1}$ with $0<\epsilon\ll 1$ (because $\omega_1\gg1$). That means that if $y_t$ is a slow variable for the time scale $t$ it is also slow for the faster time scale $\omega_1 t$.

%-------------------------------------------------------------------------------------------------%
%-------------------------------------------------------------------------------------------------%
% Section
%-------------------------------------------------------------------------------------------------%
%-------------------------------------------------------------------------------------------------%
\section{Equations governing the stochastic slow dynamics}\label{sec:SSF}

Following Roberts and Spanos~\cite{Roberts1986} the stochastic averaging method~\cite{stratonovich1967topics,Khasminskii1966}, whose general formulation is recalled in Appendix~\ref{app:3}, is used to obtain the stochastic slow dynamics of Eq.~\eqref{eq:modclarstoch2}. To apply the method, the response process $(p_t,\dot{p}_t)$ needs to be transformed into a pair of slowly varying processes. To achieve that, as in classical averaging methods of deterministic systems, an amplitude-phase representation is used imposing
\begin{subequations}
\label{eq:IllStocAveClar1}
\begin{align}
p_t &=x_t \cos\left(t+\varphi_t\right)\label{eq:IllStocAveClar1a}\\
\dot{p}_t &=-x_t \sin\left(t+\varphi_t\right)\label{eq:IllStocAveClar1b}
\end{align}
\end{subequations}

The desired form of Eq.~\eqref{eq:IllStocAveClar1b} requires that
\begin{equation}
\dot{x}_t\cos\phi_t -x_t\dot{\varphi}_t\sin\phi_t=0,
\end{equation}
with
$
\phi_t=t+\varphi_t
$
which yields
\begin{equation}
\dot{\varphi}_t=\dfrac{\dot{x}_t}{x_t}\dfrac{\cos\phi_t}{\sin\phi_t}
\quad \text{and} \quad \dot{x}_t=x_t \dot{\varphi}_t \dfrac{\sin\phi_t}{\cos\phi_t}.
\label{eq:IllStocAve3}
\end{equation}
Then differentiation of Eq.~\eqref{eq:IllStocAveClar1b} leads to
\begin{equation}
\ddot{p}_t= -x_t \cos\phi_t-\dot{x}_t\sin\phi_t -x_t\dot{\varphi} \cos\phi_t.
\label{eq:IllStocAve4}
\end{equation}
Finally, the substitution of \eqref{eq:IllStocAveClar1} and \eqref{eq:IllStocAve4} into \eqref{eq:modclarstoch2} and the use of \eqref{eq:IllStocAve3} yields
\begin{subequations}
\label{eq:ampphaseCar1}
\begin{align}
\dot{x}_t &=  h\left(x_t\cos\phi _t,-x_t\sin\phi_t,y_t\right) \sin\phi_t
-\nu\xi_t \sin\phi_t\label{eq:ampphaseCar1a}\\
\dot{\varphi}_t &= 
h\left(x_t\cos\phi_t ,-x_t\sin\phi_t,y_t\right) \frac{\cos\phi_t}{x_t}
-\nu\xi_t \frac{\cos\phi_t}{x_t}\label{eq:ampphaseCar1b}
\end{align}
\end{subequations}
which has a similar form as \eqref{eq:stochave1} in Appendix~\ref{app:3} with
\begin{equation}
\begin{split}
{\bf x}_t &= (x_t,\varphi_t)^T, \qquad \boldsymbol{\eta}_t = (\xi_t ,\xi_t )^T,\\
{\bf f}({\bf x}_t,t)&=
\begin{pmatrix}
  h\left(x_t\cos\phi_t ,-x_t\sin\phi_t,y_t\right) \sin\phi_t \\
\dfrac{1}{x_t}  h\left(x_t\cos\phi_t,-x_t\sin\phi_t,y_t\right) \cos\phi_t
\end{pmatrix},\\
{\bf g}({\bf x}_t,t)&=
\begin{pmatrix}
-\nu\sin\phi_t& 0\\
0 & -\nu\dfrac{\cos\phi_t}{x_t}
\end{pmatrix}
\end{split}
\end{equation}
with $()^T$ the transpose operator.

Following the approach described in Appendix~\ref{app:3}, the drift vector $\bf m$ (see Eq.~\eqref{eq:stochavem}) is first computed. The first term 
$
T^{\text{av}}
\left\lbrace
\bf f
\right\rbrace=\left(F(x_t,y_t),G(x_t,y_t)\right)^T,
$
with
\begin{align}
F(x_t,y_t)
&=\left(
\frac{F_1}{\omega_1}\zeta \frac{3 (y_t+\hat{\gamma}^{\text{st}}) -1 }{4 (y_t+\hat{\gamma}^{\text{st}})^{1/2}}
-\frac{\alpha_1}{2}\right)x_t\nonumber\\
&+\frac{F_1}{\omega_1}\zeta\frac{3  \left(\hat{\gamma}^{\text{st}} +y_t+1\right)}{128 \left(\hat{\gamma}^{\text{st}}+y_t\right){}^{5/2}}x_t^3 \label{eqF1}\\
G(x_t,y_t)
&=0,\label{eqG1}
\end{align}
corresponds to the classical (deterministic) Bogoliubov-Krylov averaging of the vector function $\bf f$, i.e. the time average is performed assuming that $x_t$ and $\phi_t$ are slow variables with respect to the unitary eigenfrequency of the considered dimensionless simplified reed instrument model~\eqref{eq:modclarstoch2} which therefore corresponds to a period equal to $2\pi$.

Then, the second part of the drift vector $\bf m$ (see again Eq.~\eqref{eq:stochavem}) is determined as
\begin{align}
&T^{\text{av}}
\left\lbrace
\int_{-\infty}^0
\EE{
\left(\frac{\partial{(\bf g \boldsymbol{\eta})}}{\partial{\bf x} } \right)_t ({\bf g} \boldsymbol{\eta})_{t+\tau}}d\tau
\right\rbrace\nonumber\\
 &=\frac{1}{2 \pi}\int_{0}^{2\pi}
\int_{-\infty}^0
\EE{
\left(\frac{\partial{(\bf g \boldsymbol{\eta})}}{\partial{\bf x} } \right)_t ({\bf g} \boldsymbol{\eta})_{t+\tau}}d\tau
dt\nonumber\\
% &=
%\int_{-\infty}^0
%\frac{1}{2 \pi}
%\int_{t_0}^{t_0+2\pi}
%\EE{
%\left(\frac{\partial{(\bf g \boldsymbol{\eta})}}{\partial{\bf x} } \right)_t ({\bf g} \boldsymbol{\eta})_{t+\tau}}dtd\tau\nonumber\\
 &=
 \frac{1}{2 \pi}
\int_{-\infty}^0
\int_{0}^{2\pi}\nonumber\\
&\qquad\qquad\delta(\tau)
\begin{pmatrix}
\frac{\nu^2\cos\phi_t\cos\phi_{t+\tau}}{x_t}\\
-\nu^2\frac{\sin\phi_t\cos\phi_{t+\tau}+\cos\phi_t\sin\phi_{t+\tau}}{x_t^2}
\end{pmatrix}
dtd\tau\nonumber\\
&=
\begin{pmatrix}
\frac{\nu^2}{4x_t}\\
0
\end{pmatrix}
\end{align}
in which the symbol $\EE{}$ disappeared due to definitions~\eqref{eq:white1} and one used the fact that 
$
\int_{-\infty}^0 \delta(\tau) \cos(\tau)d\tau=\frac{1}{2}\int_{-\infty}^{+\infty} \delta(\tau) \cos(\tau)d\tau=\frac{1}{2}
$.
The final expression of the drift vector is therefore
\begin{equation}
{\bf m}=
\begin{pmatrix}
F(x_t,y_t)+\dfrac{\nu^2}{4 x_t}\\
0
\end{pmatrix}.
\label{eq:exm}
\end{equation}

Now the expression of the diffusion matrix $\boldsymbol{\sigma}$ is determined from \eqref{eq:stochavesig} starting by computing
\begin{align}
& T^{\text{av}}
\left\lbrace
\int_{-\infty}^{+\infty}
\EE{
({\bf g} \boldsymbol{\eta})_{t}
({\bf g} \boldsymbol{\eta})_{t+\tau}^T}d\tau
\right\rbrace\nonumber\\
&=\frac{1}{2 \pi}
\int_{0}^{2\pi}
\int_{-\infty}^{+\infty} 
\EE{
({\bf g} \boldsymbol{\eta})_{t}
({\bf g} \boldsymbol{\eta})_{t+\tau}^T}d\tau dt\nonumber\\
%&=
%\int_{-\infty}^{+\infty} 
%\frac{1}{2 \pi}
%\int_{t_0}^{t_0+2\pi}
%\EE{
%({\bf g} \boldsymbol{\eta})_{t}
%({\bf g} \boldsymbol{\eta})_{t+\tau}^T}dt d\tau\nonumber\\
&=
\int_{-\infty}^{+\infty} 
\begin{pmatrix}
\frac{\nu^2\delta(\tau)\cos(\tau)}{2}  & -\frac{\nu^2\delta(\tau)\sin(\tau)}{2 x_t}\\
\frac{\nu^2\delta(\tau)\sin(\tau)}{2 x_t} & \frac{\nu^2\delta(\tau)\cos(\tau)}{2 x_t^2}
\end{pmatrix}
d\tau\nonumber\\
&=
\begin{pmatrix}
\dfrac{\nu^2}{2}  & 0\\
0 & \dfrac{\nu^2}{2 x_t^2}
\end{pmatrix}.
\label{eq:calcsig1}
\end{align}
in which again definitions \eqref{eq:white1} have been used. From~\eqref{eq:calcsig1} a possible solution of \eqref{eq:stochavesig} is
\begin{equation}
\boldsymbol{\sigma}=
\begin{pmatrix}
\dfrac{\nu}{\sqrt{2}}& 0\\
0 & \dfrac{1}{x_t}\dfrac{\nu}{\sqrt{2}}
\end{pmatrix}.
\label{eq:exsig}
\end{equation}
Therefore, according to~\eqref{eq:stochave2} the equations governing the stochastic slow dynamics of \eqref{eq:modclarstoch2} are
\begin{subequations}
\label{eq:ampphaseClarMoyAnn}
\begin{align}
d x_t &=\left(F(x_t,y_t)+\frac{\nu^2}{4 x_t}\right)dt+\dfrac{\nu}{\sqrt{2}}dW_t\label{eq:ampphaseClarMoya}\\
d \varphi_t &=\frac{1}{x_t}\dfrac{\nu}{\sqrt{2}} dW_t.
\end{align}
\end{subequations}
in which the time evolution of the amplitude $x_t$ is uncoupled from that of the phase $\varphi_t$.

Assuming a weak noise level (i.e. $\nu\ll1$) and therefore neglecting the term $\frac{\nu^2}{4 x_t}$ in \eqref{eq:ampphaseClarMoya} and denoting $\sigma=\frac{\nu}{\sqrt{2}}$ we obtain the following Itô stochastic differential equation for the amplitude of the mouth pressure~$p_t$
\begin{equation}
dx_t=
F(x_t,y_t)dt + \sigma dW_t.
\label{eq:VdP1Moy3Stocha}
\end{equation}

The scaling and time shift properties of Wiener process (see~\cite{bernt2003}, Chaps.~7) allow us to write that\\
$
\frac{1}{\sqrt{\epsilon}}
\left(
W_{\epsilon t +y_0}-W_{y_0}
\right)
\sim
\frac{1}{\sqrt{\epsilon}}
W_{\epsilon t}
\sim
W_{t}
$
and therefore
\begin{equation}
\frac{1}{\sqrt{\epsilon}}
dW_{y}
\sim
dW_{t}.
\label{eq:scalty}
\end{equation}

Finally, from \eqref{eq:VdP1Moy3Stocha} and \eqref{eq:scalty} we obtain the final form of the slow dynamics as
\begin{equation}
dx_y=\frac{1}{\epsilon}F(x_y,y)dy + \frac{\sigma}{\sqrt{\epsilon}} dW_y.
\label{eq:VdP1Moy3Stochay}
\end{equation}
where $x_t$ and $y_t$ are now denoted $x_y$ and $y$ respectively to emphasize that Eq.~\eqref{eq:VdP1Moy3Stochay} is a non-autonomous one-dimensional Itô stochastic differential equation with respect to $x_y$ depending on $y$ which acts as the time variable.

%-------------------------------------------------------------------------------------------------%
%-------------------------------------------------------------------------------------------------%
% Section
%-------------------------------------------------------------------------------------------------%
%-------------------------------------------------------------------------------------------------%
\section{Analytical expression of the dynamic pitchfork bifurcation points of the averaged system}
\label{sec:Analy}

%In this section the analytical expression of the dynamic Hopf bifurcation point of Eq.~\eqref{eq:modclarstoch2} is obtained from the analysis of \eqref{eq:VdP1Moy3Stochay}.

The supercritical Hopf bifurcation at $y=0$ for the initial (non averaged) system \eqref{eq:modclarstoch2} becomes a supercritical pitchfork bifurcation for Eq.~\eqref{eq:VdP1Moy3Stocha}. Therefore, if the averaged system~\eqref{eq:VdP1Moy3Stocha} is a good approximation of \eqref{eq:modclarstoch2}, the dynamic Hopf bifurcation point of Eq.~\eqref{eq:modclarstoch2} is very close to the dynamic pitchfork bifurcation point of \eqref{eq:VdP1Moy3Stocha}. The analytical expression of the latter is obtained in this section in both deterministic and stochastic cases.

As stated by Stocks et al.~\cite{Stocks1989}, the dynamic bifurcation point\footnote{Sometimes called \textit{exit value} in the literature.} has two main definitions: it is (1) the value of the bifurcation parameter $y$ when $\sqrt{\EE{x_{y}^2}}$ crosses a predefined threshold $x_{\text{th}}^2$ (here the initial value is chosen) or (2) the value of the bifurcation parameter $y$ at which, on average, the random variable $x_{y}^2$ crosses the same given threshold $x_{\text{th}}^2$. In general, these two averaging procedures will lead to two slightly different results. In the paper the first definition is chosen and is clearly recalled below. Other definitions exist (see e.g. \cite{Baesens1991Noise} and \cite{BergeotNLD2012b} for a clarinet-like system).

\begin{defn}[Dynamic pitchfork bifurcation point]\label{def:ptbifdyn}
In the case of a pitchfork bifurcation and in a deterministic framework, the dynamic bifurcation point is defined as the value of $y$ for which $x_y$ exceeds its initial value $x_{y_0}$ (in absolute value). In a stochastic framework, as in this work, the expected value of the squared amplitude is considered and the bifurcation point, denoted $\hat{y}^{\text{dyn}}$, is defined as the value of $y_t$ for which  $\sqrt{\EE{x_{y}^2}}$ exceeds the initial value $x_{y_0}$. The dynamic bifurcation point is larger than the static bifurcation and the difference between them is called \textbf{bifurcation delay}.
%denoted $\hat{y}^{\text{dyn}}$,
\end{defn}

%Consequently, the first step is to compute $\EE{x_{y}^2}$, this is performed in the next section. 

%_____________________________Subsection_____________________________%
%\subsection{Expected value of the squared amplitude}\label{sec:EVA}
\subsection{Solution of the linearized averaged system}\label{sec:EVA}

The linearized version of \eqref{eq:VdP1Moy3Stochay} with respect to $x_y$ around 0 is considered
\begin{equation}
dx_y=
\frac{1}{\epsilon}a(y)x_ydy + \frac{\sigma}{\sqrt{\epsilon}} dW_y
\label{eq:VdP1Moy3StochayLin}
\end{equation}
where, from Eq.~\eqref{eqF1}, one has
\begin{equation}
a(y)=\frac{\partial F}{\partial x_y}(0,y)=
\frac{F_1}{\omega_1}\zeta \frac{3 (y+\hat{\gamma}^{\text{st}}) -1 }{4 (y+\hat{\gamma}^{\text{st}})^{1/2}}
-\frac{\alpha_1}{2}.
\label{eq:defa}
\end{equation}

In order to solve \eqref{eq:VdP1Moy3StochayLin}, the deterministic differential equation associated to the latter is first considered as
\begin{equation}
\frac{d x_y}{d y}=\frac{1}{\epsilon}a(y)x_y
\label{eq:VdP1Moy3Det}
\end{equation}
whose solution, denoted $x^\text{det}(y)$, is
\begin{equation}
x^\text{det}(y) = x_{y_0} e^{\frac{1}{\epsilon}(A(y)-A(y_0))}
\label{eq:VdP1Moy3DetSol}
\end{equation}
where
\begin{equation}
%A(y)&=\int_0^y a(y)dy\nonumber\\
A(y)=\frac{\zeta  F_1 (\hat{\gamma}^{\text{st}} +y-1) \sqrt{\hat{\gamma}^{\text{st}} +y}}{2\omega _1}-\frac{\alpha_1}{2} y.
\label{eq:defA}
\end{equation}
is the antiderivative of $a(y)$.

Following a well-known method for solving stochastic differential equations, we apply the Itô formula (see Eq.~\eqref{eq:ItoFormula} in Appendix~\ref{app:ItoFormula}) to the function $f(x_y,y)=x_{y} e^{-\frac{1}{\epsilon}(A(y)-A(y_0))}$ which is constant for the deterministic Eq.~\eqref{eq:VdP1Moy3Det} (indeed through Eq.~\eqref{eq:VdP1Moy3DetSol} one has $f(x^\text{det}(y),y)=x_{y_0}$). Using the differential of the product rule, and noting that in the Itô formula the term $\frac{\partial f}{\partial y} +\frac{1}{\epsilon}a(y)x_y\frac{\partial f}{\partial x}$ vanishes, we obtain
%\begin{align}
%d\left(x_{y}e^{-\frac{1}{\epsilon}(A(y)-A(y_0))}\right)&=
%\Bigg(-x_y \frac{1}{\epsilon} \frac{d A(y)}{dy} e^{-\frac{1}{\epsilon}(A(y)-A(y_0))}\nonumber\\
%&+x_y\frac{1}{\epsilon}
%a(y) e^{-\frac{1}{\epsilon}(A(y)-A(y_0))}\Bigg)dy\nonumber\\
%&+\frac{\sigma}{\sqrt{\epsilon}}  e^{-\frac{1}{\epsilon}(A(y)-A(y_0))} dW_y\nonumber\\
%&=
%\frac{\sigma}{\sqrt{\epsilon}}  e^{-\frac{1}{\epsilon}(A(y)-A(y_0))} dW_y
%\label{eq:d1}
%\end{align}
\begin{equation}
df(x_y,y)=
\frac{\sigma}{\sqrt{\epsilon}}  e^{-\frac{1}{\epsilon}(A(y)-A(y_0))} dW_y
\label{eq:d1}
\end{equation}
and integrating \eqref{eq:d1} from $y_0$ to $y$ yields the following solution of Eq.~\eqref{eq:VdP1Moy3StochayLin}
\begin{equation}
x_y =x^\text{det}(y) +
%x_{y_0} e^{\frac{1}{\epsilon}(A(y)-A(y_0))}+
\frac{\sigma}{\sqrt{\epsilon}}e^{\frac{1}{\epsilon}A(y)}\int_{y_0}^{y}e^{-\frac{1}{\epsilon}A(y')}dW_{y'}.
\label{eq:solPitchGeby2}
\end{equation}
Note that the first and second terms of the right-hand side of \eqref{eq:solPitchGeby2} are the deterministic and stochastic parts of $x_y$ respectively. The integral in the second term is called an Itô integral and, for a given function $g$, has the following property $\EE{\int g dW_y}=0$ assuming some properties for the function $g$ (see \cite{bernt2003}, Chap.~3) that are respected by $e^{-\frac{1}{\epsilon}A(y)}$. Therefore we have
\begin{equation}
\EE{\int_{y_0}^{y}e^{-\frac{1}{\epsilon}A(y)}dW_y}=0.
\label{eq:propW2}
\end{equation}
That means that the expected value of $x_y$ is the solution of the associated deterministic equation~\eqref{eq:VdP1Moy3Det}, i.e.
$
\EE{x_y}=x^\text{det}(y)
%=\EE{x_{y_0} e^{\frac{1}{\epsilon}(A(y)-A(y_0))}}
%=x_{y_0} e^{\frac{1}{\epsilon}(A(y)-A(y_0))}.
$.

Now $\EE{x_{y}^2}$ is computed from \eqref{eq:solPitchGeby2}, that yields
\begin{align}
\EE{x_{y}^2}&=\EE{(x^\text{det}(y))^2}\nonumber\\
&+\EE{\frac{\sigma^2}{\epsilon}e^{\frac{2}{\epsilon}A(y)}\left(\int_{y_0}^{y}e^{-\frac{1}{\epsilon}A(y')}dW_{y'}\right)^2}\nonumber\\
&+\EE{2 \, x^\text{det}(y) \frac{\sigma}{\sqrt{\epsilon}} e^{\frac{1}{\epsilon}A(y)}\ \int_{y_0}^{y}e^{-\frac{1}{\epsilon}A(y')}dW_{y'}}.
\label{eq:EVSA1}
\end{align}
Using again the property of the Itô integral which states that $\EE{\int g dW_y}=0$, the third term in the right-hand side of \eqref{eq:EVSA1} is equal to zero. The second term is simplified using the \textit{Itô isometry} (see Corollary 3.1.7 in~\cite{bernt2003}, Chap.~3) which states that
$
\EE{\left(\int g dW_y\right)^2}
=
\EE{\int g^2 dy}
$
assuming again some properties for the function $g$ that are resected by $e^{-\frac{1}{\epsilon}A(y)}$. Therefore we obtain the final expression of the expected value of the squared amplitude
\begin{equation}
\EE{x_{y}^2}=\mathcal{D}(y)+\mathcal{S}(y)
\label{eq:EVSA2}
\end{equation}
where
\begin{subequations}
\label{eq:EVSA2b}
\begin{align}
\mathcal{D}(y)&=(x^\text{det}(y))^2\\
\mathcal{S}(y)&=\frac{\sigma^2}{\epsilon}e^{\frac{2}{\epsilon}A(y)}\int_{y_0}^{y}e^{-\frac{2}{\epsilon}A(y)}dy\label{eq:EVSA2bb}
\end{align}
\end{subequations}
are respectively the deterministic and the stochastic parts of the expected value of the squared amplitude.

From \eqref{eq:EVSA2} and according to \cite{BGProb2002}, three regimes can be distinguished for a system such as \eqref{eq:VdP1Moy3Stochay} undergoing a dynamic pitchfork bifurcation:

\begin{itemize}
\item \textit{Regime I:} $\mathcal{S}(y)\ll \mathcal{D}(y)$. In this case the noise level is so small that it can be neglected and the problem is identical to the deterministic case \eqref{eq:VdP1Moy3Det} which undergoes the greatest possible bifurcation delay.
\item \textit{Regime II:} $\mathcal{S}(y)\gg\mathcal{D}(y)$ with a noise level not too high. The noise can no longer be neglected but it remains small enough for linearized model~\eqref{eq:VdP1Moy3StochayLin} to remain valid. A bifurcation delay still exists but it is reduced compared to Regime I.
\item \textit{Regime III:} $\mathcal{S}(y)\gg\mathcal{D}(y)$ with a high noise level. In this situation the behavior of the system is dominated by noise and the trajectory of $x_y$ leaves the neighborhood of zero before the static bifurcation point $y=0$ is reached. In such a case the linear approximation is therefore not valid anymore and, as highlighted by Berglung and Gentz~\cite{BGProb2002}, the notion of bifurcation delay becomes meaningless.
\end{itemize}

The domains of existence of each regime are given explicitly in Sect.~\ref{sec:opreg}.

In the two following parts, we concentrate on Regimes~I and II, giving the analytical expression of the dynamic bifurcation points considering each regime separately.

%\begin{figure}[t!]
%	\centering
%	\subfigure[]{\includegraphics[width=1\columnwidth]{pProbaSig10m4.eps}}
%	\subfigure[]{\includegraphics[width=1\columnwidth]{pProbaSig10m6.eps}}
%	\caption{}
%\label{fig:PDF}
%\end{figure}

%_____________________________Subsection_____________________________%
\subsection{Dynamic bifurcation point in the deterministic case}\label{sec:AnaDet}

In this section we assume that the system is in Regime~I and therefore $\EE{x_{y}^2}=\mathcal{D}(y)=(x^\text{det}(y))^2$ (with $x^\text{det}(y)$ given by Eq.~\eqref{eq:VdP1Moy3DetSol}). %In this case, the system is deterministic and we have $\EE{x_{y}^2}=x_{y}^2$. Consequently $x_{y}=\sqrt{\mathcal{D}(y)}$ which is given by Eq.~\eqref{eq:VdP1Moy3DetSol}.
Consequently, from Definition~\ref{def:ptbifdyn}, the deterministic dynamic bifurcation point, denoted $\hat{y}^{\text{dyn}}_{\text{det}}$, is a solution of
$
x_{y_0} = x_{y_0} e^{\frac{1}{\epsilon}(A(y)-A(y_0))}
$
and therefore of
\begin{equation}
A(y)=A(y_0).
\label{eq:AA0}
\end{equation}

%The expressions $a(y)$ ans $A(y)$ are easily deduced from \eqref{eqF2}, \eqref{eq:defa} and \eqref{eq:defA}, they are
%\begin{equation}
%a(y)=\frac{F_1}{\omega_1}\zeta \frac{3 (y+\hat{\gamma}^{\text{st}}) -1 }{4 (y+\hat{\gamma}^{\text{st}})^{1/2}}
%-\frac{\alpha_1}{2}
%\end{equation}
%and
%\begin{equation}
%A(y)=\frac{F_1}{2\omega_1}\zeta 
%\left(
%(y+\hat{\gamma}^{\text{st}})^{3/2}-(y+\hat{\gamma}^{\text{st}})^{1/2}
%\right)
%-\frac{\alpha_1}{2}y.
%\label{eqExpA}
%\end{equation}

\begin{figure}[t!]
	\centering
	\includegraphics[width=1\columnwidth]{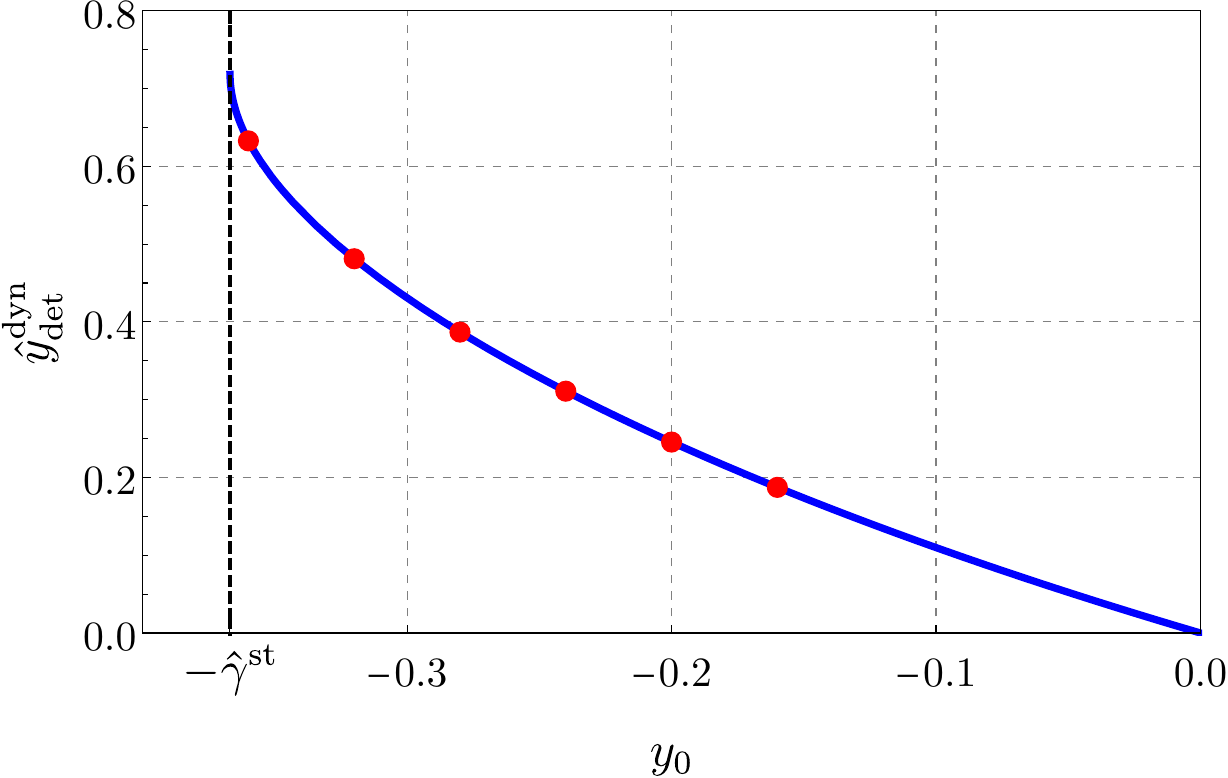}
	\caption{Deterministic dynamic bifurcation point $\hat{y}^{\text{dyn}}_{\text{det}}$, given by Eq.~\eqref{eq:dyndetbifpt1}, as a function of the initial condition $y_0$. The opposite of the static bifurcation point $\hat{\gamma}^{\text{st}}$, given by Eq.~\eqref{eq:statbifpt}, is depicted by a vertical dashed line. The red points correspond to the initial conditions used in Fig.~\ref{fig:pDBPDetNum}. The set of parameters \eqref{eq:param1} is used.}
	\label{fig:pDBPDet}
	%param = {\[Alpha]1 -> 0.02`100, \[Omega]1 -> 1000`100,   F1 -> 1200`100, \[Epsilon] -> 0.002`100, \[Zeta] -> 0.2`100};
\end{figure}

From \eqref{eq:defA}, Eq.~\eqref{eq:AA0} is solved (details are given in Appendix~\ref{app:2}). It is shown that, in addition to the trivial solution $y=y_0$, Eq.~\eqref{eq:AA0} has two other solutions. The one that corresponds to the deterministic dynamic bifurcation point is
\begin{equation}
\hat{y}^{\text{dyn}}_{\text{det}}=X_2^2-\hat{\gamma}^{\text{st}},
\label{eq:dyndetbifpt1}
\end{equation}
where $X_2$ is given by Eq.~\eqref{eq:X2}. We will see in Fig.~\ref{fig:pDBPDet} that the maximum of $\hat{y}^{\text{dyn}}_{\text{det}}$ is obtained for $y_0=-\hat{\gamma}^{\text{st}}$ and in this case Eq.~\eqref{eq:dyndetbifpt1} reduces to 
\begin{equation}
\hat{y}^{\text{dyn}}_{\text{det}}=\frac{\left(\alpha _1 \omega _1+\sqrt{\alpha _1^2 \omega _1^2+4 \zeta ^2 F_1^2}\right){}^2}{4 \zeta ^2 F_1^2}-\hat{\gamma}^{\text{st}}
\label{eq:dyndetbifpt1b}
\end{equation}
It is important to be aware that even if $\hat{y}^{\text{dyn}}_{\text{det}}$ has a finite value for $y_0=-\hat{\gamma}^{\text{st}}$, the function $F(x,y) $ in Eq.~\eqref{eq:VdP1Moy3Stochay}  diverge at $y=-\hat{\gamma}^{\text{st}}$. Therefore, Eq.~\eqref{eq:dyndetbifpt1b} must be understood as a limit value for $y_0$ very close to $-\hat{\gamma}^{\text{st}}$ (from above). However, for values of $y$ very close to $-\hat{\gamma}^{\text{st}}$, the linearization of Eq.~\eqref{eq:VdP1Moy3Stochay} with respect to $x$ (leading to Eq.~\eqref{eq:VdP1Moy3StochayLin}) is valid if $x$ is very close to zero. In other words, Eq.~\eqref{eq:dyndetbifpt1b} is valid to predict the deterministic bifurcation point of a numerical simulation of  Eq.~\eqref{eq:VdP1Moy3Stochay} only if $y_0$ and $x_{y_0}$ are very close to $-\hat{\gamma}^{\text{st}}$ and 0 respectively. This corresponds to the situation where the musician begins to blow gently into the instrument.

In the lossless case, i.e. $\alpha_1=0$ and $\hat{\gamma}^{\text{st}}=\frac{1}{3}$, Eq.~\eqref{eq:dyndetbifpt1} becomes
\begin{equation}
\hat{y}^{\text{dyn}}_{\text{det}}=\frac{1}{2}\left(1-y_0-\sqrt{1+y_0(2-3 y_0)}\right)
\label{eq:dyndetbifpt2}
\end{equation}
which depends only on the initial value $y_0$. Note first that Eq.~\eqref{eq:dyndetbifpt2} yields $\hat{y}^{\text{dyn}}_{\text{det}}=2/3$ if $y_0=-\hat{\gamma}^{\text{st}}=-\frac{1}{3}$. 

This is the largest value of the dynamic bifurcation point. From a physical point of view this means that the oscillations emerge when the blowing pressure $\gamma=\hat{\gamma}^{\text{st}} +\hat{y}^{\text{dyn}}_{\text{det}}=1$. However this value is known to be the limit of validity of the clarinet model since the reed channel becomes completely closed (an effect not taken into account in the model considered in the work). Therefore, in the lossless case a linear increase of the blowing pressure from an arbitrary small value gives a scenario where the clarinet never plays. This remark can be extended, as shown below, to situations in which damping is taken into account (i.e. $\alpha_1> 0$).

Then, in cases with and without damping, $\hat{y}^{\text{dyn}}_{\text{det}}$ does not depend on the slope $\epsilon$ which may seem counterintuitive.

\begin{figure}[t!]
	\centering
	\includegraphics[width=1\columnwidth]{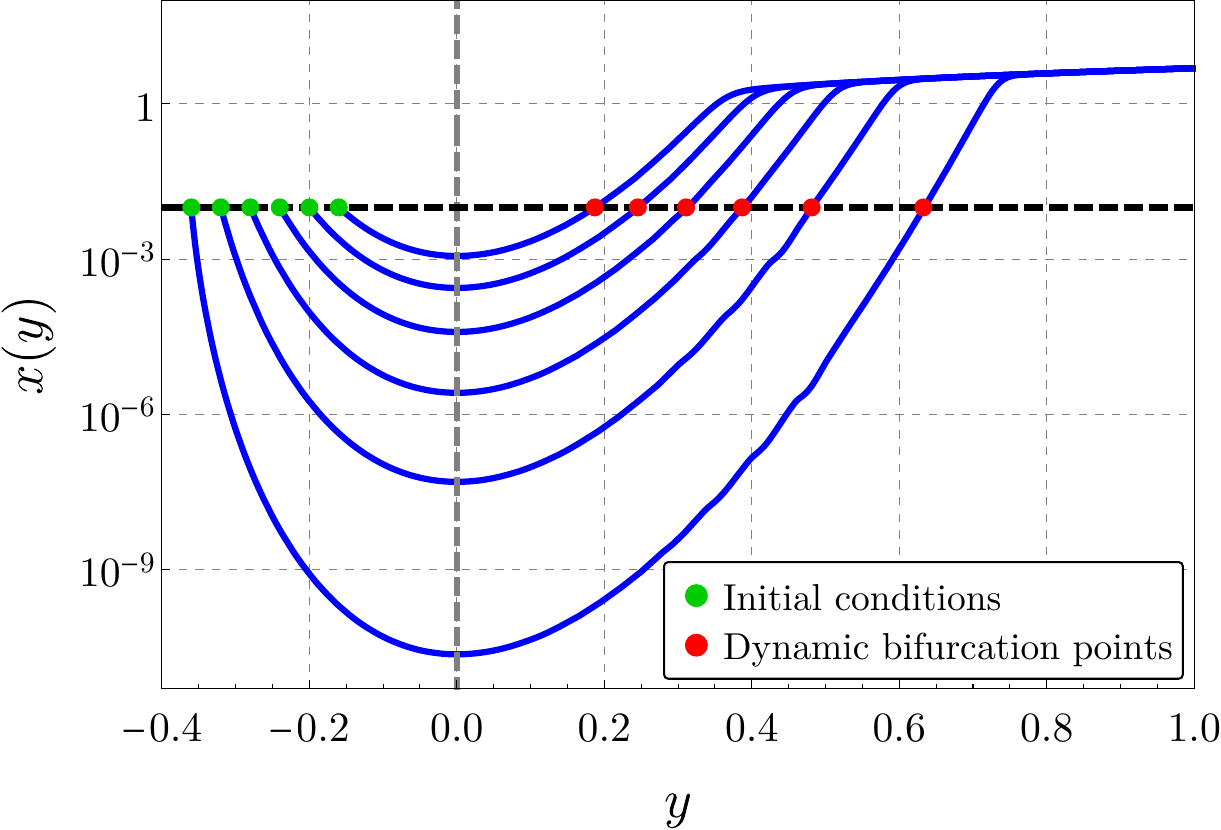}
	\caption{Numerical simulations of the deterministic differential equation associated to Eq.~\eqref{eq:VdP1Moy3Stochay} using a logarithm scale for the vertical axis.  The set of parameters \eqref{eq:param1} is used with, in addition, $x(y_0)=0.01$ (the latter is depicted by a horizontal dashed line) and $y_0=-0.36,-0.32,\dots,-0.16$. The static bifurcation point $\hat{y}^{\text{st}}=0$ is depicted by a vertical gray dashed line. The dynamic bifurcation point $\hat{y}^{\text{dyn}}_{\text{det}}$ is defined for a given initial condition by the value of $y$ for which $|x(y)|=|x(y_0)|$, which corresponds graphically to the intersection between the blue curves and the horizontal dashed line. The green and red points are used respectively to highlight the considered initial conditions and the corresponding dynamic bifurcation points predicted by Eq.~\eqref{eq:dyndetbifpt1}.}
	\label{fig:pDBPDetNum}
	%param = {\[Alpha]1 -> 0.02`100, \[Omega]1 -> 1000`100,   F1 -> 1200`100, \[Epsilon] -> 0.002`100, \[Zeta] -> 0.2`100};
\end{figure}

The deterministic dynamic bifurcation point $\hat{y}^{\text{dyn}}_{\text{det}}$ , given by Eq.~\eqref{eq:dyndetbifpt1}, is plotted in Fig.~\ref{fig:pDBPDet} as a function of the initial condition $y_0$ with the following set of parameters:
\begin{equation}
\begin{gathered}
\epsilon=0.002, \quad \omega_1=1000 \; \text{rad}\cdot \text{s}^{-1}, \quad \alpha_1=0.02,\\
F_1=1200 \; \text{s}^{-1} \quad \text{and} \quad \zeta=0.2.
\end{gathered}
\label{eq:param1}
\end{equation}
The figure shows that $\hat{y}^{\text{dyn}}_{\text{det}}$ is a decreasing function with respect to $y_0$ and, as previously mentioned, the maximum is obtained for $y_0=-\hat{\gamma}^{\text{st}}$. Moreover, the red points correspond to the initial conditions used in Fig.~\ref{fig:pDBPDetNum}. In the latter, numerical simulations of the deterministic differential deterministic differential equation associated to Eq.~\eqref{eq:VdP1Moy3Stochay} are shown with a logarithmic scale for the vertical axis. The same parameters are as in Fig.~\ref{fig:pDBPDet} and the initial values are: $x(y_0)=0.01$ and $y_0=-0.36,-0.32,\dots,-0.16$. The dynamic bifurcation point is defined for a given initial condition by the value of $y$ for which $|x(y)|=|x(y_0)|$, which corresponds graphically to the intersection between the blue curves and the horizontal dashed line. The minimum is reached at the static bifurcation point $\hat{y}^{\text{st}}=0$ (depicted by a vertical dashed gray line). One can see that the smaller $y_0$, the smaller the minimum of the trajectory $x(y)$ too and consequently the longer the path to travel before reaching $x(y_0)$ leading to larger values of the deterministic dynamic bifurcation point $\hat{y}^{\text{dyn}}_{\text{det}}$. The green and red points are used respectively to highlight the considered initial conditions and the corresponding dynamic bifurcation points predicted by Eq.~\eqref{eq:dyndetbifpt1}. This shows that the theoretical results presented in Fig.~\ref{fig:pDBPDet} predict the deterministic bifurcation points measured on numerical simulations.

\begin{figure}[t!]
	\centering
	\includegraphics[width=1\columnwidth]{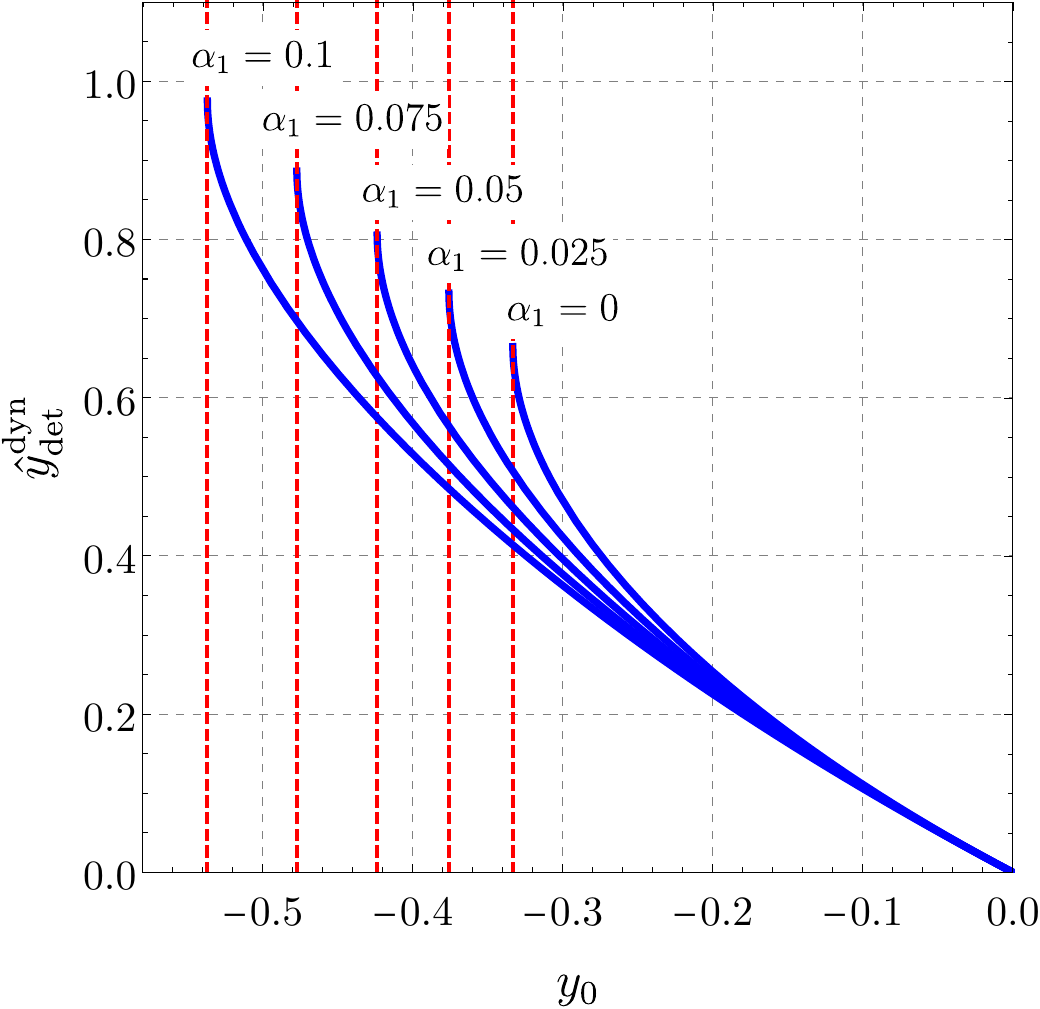}
	\caption{The deterministic bifurcation point $\hat{y}^{\text{dyn}}_{\text{det}}$ given by Eq.~\eqref{eq:dyndetbifpt1} as a function of the initial condition $y_0$ for five values of the damping coefficient, i.e. $\alpha_1=0,0.025\dots,0.1$ with $\omega_1=1000$ rad$\cdot$s$^{-1}$, $F_1=1200$ s$^{-1}$ and $\zeta=0.2$. Red dashed lines indicate the value of $-\hat{\gamma}^{\text{st}}$ for each value of $\alpha_n$.}
	\label{p3DDet1b}
\end{figure}

\begin{figure}[t!]
	\centering
	\includegraphics[width=1\columnwidth]{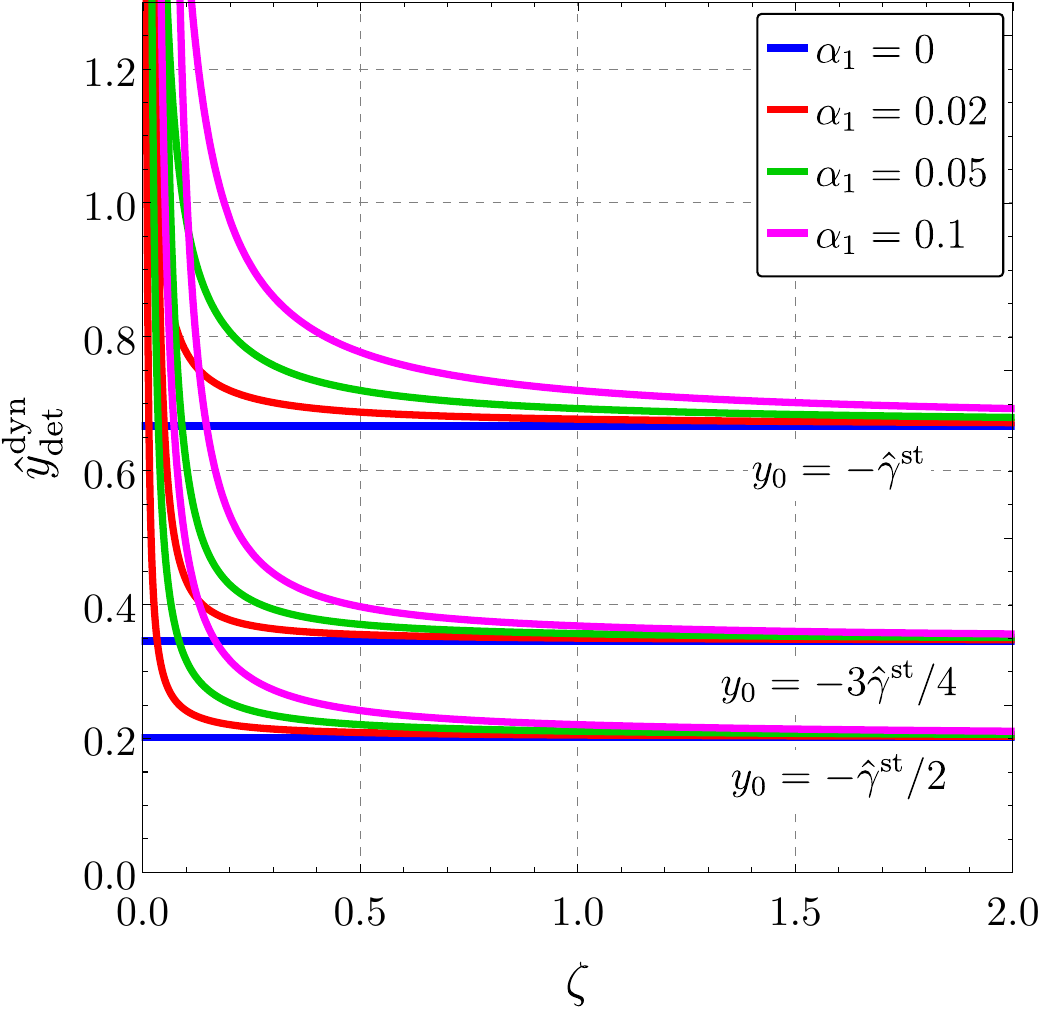}%\label{fig:pDBPDet2a}
	\caption{The deterministic bifurcation point $\hat{y}^{\text{dyn}}_{\text{det}}$ given by Eq.~\eqref{eq:dyndetbifpt1} as a function of the parameter $\zeta$ for four of the damping coefficient ($\alpha_1=0,0.01,0.02$ and $0.05$), for three value of the initial condition ($y_0=-\hat{\gamma}^\text{st},-3\hat{\gamma}^\text{st}/4$ and $-\hat{\gamma}^\text{st}/2$) with $\omega_1=1000$ rad$\cdot$s$^{-1}$ and $F_1=1200$ s$^{-1}$.}
	\label{p3DDet1c}
\end{figure}

In the lossless case the expression of the deterministic dynamic bifurcation point $\hat{y}^{\text{dyn}}_{\text{det}}$ is given by \eqref{eq:dyndetbifpt2} which depends only on the initial value $y_0$. The deterministic bifurcation point $\hat{y}^{\text{dyn}}_{\text{det}}$ given by Eq.~\eqref{eq:dyndetbifpt1} is plotted as a function of the initial condition $y_0$ for five values of the damping coefficient in Fig.~\ref{p3DDet1b} and as a function of the parameter $\zeta$ for four values of the damping coefficient $\alpha_1$ and for three values of the initial condition $y_0$ in Fig.~\ref{p3DDet1c}. The figures show that in the case with a small damping, the initial value $y_0$ remains the most influential parameter except for the smallest values of the control parameter $\zeta$ and for initial conditions $y_0$ close to the opposite of the static bifurcation point $\hat{\gamma}^\text{st}$ (see Fig.~\ref{p3DDet1c}). In general one has $0.1<\zeta<0.4$ for a clarinet, $0.25<\zeta<1$ for a saxophone and more for double-reed instruments.

As said previously, the deterministic dynamic bifurcation point corresponds to the largest possible bifurcation delay which holds when the noise can be neglected. In general, because the noise prevents the amplitude $x_y$ to have a very small value when the parameter $y$ reaches its static bifurcation value, the effect of an additive noise is to reduce the delay and to make it lose its dependence on the initial condition $y_0$. The influence of noise is studied in the next section.

%_____________________________Subsection_____________________________%
\subsection{Dynamic bifurcation point in the stochastic case}\label{sec:AnaStoch}

Now the system is assumed to evolve according to Regime II, i.e. $\EE{x_{y}^2}=\mathcal{S}(y)$. In this case, again through  Definition~\ref{def:ptbifdyn} and using Eq.~\eqref{eq:EVSA2bb}, the equation to solve in order to find the dynamic bifurcation point is
\begin{equation}
x_{y_0}^2=\frac{\sigma^2}{\epsilon}e^{\frac{2}{\epsilon}A(y)}\int_{y_0}^{y}e^{-\frac{2}{\epsilon}A(y)}dy.
\label{eq:eqDynPtStoch1}
\end{equation}

The first step is to obtain the approximate expression of the integral
\begin{equation}
I(y_0,y)=\int_{y_0}^{y}e^{-\frac{2}{\epsilon}A(y)}dy.
\end{equation}
For this purpose $A(y)$ (given by Eq.~\eqref{eq:defA}) is expanded in a second order Taylor series around 0 (the static bifurcation point), that leads to
\begin{subequations}
\begin{align}
A(y)&\approx A(0)+y \frac{d A}{dy}(0)+\frac{y^2}{2}\frac{d^2 A}{dy^2}(0)\\
&=A(0)+ya(0)+\frac{y^2}{2}a'(0)\label{eq:TSAb}
\end{align}
\end{subequations}
where the second term in the right-hand side of \eqref{eq:TSAb} vanishes by definition\footnote{Indeed, at the bifurcation one has $a(0)=\frac{\partial F}{\partial x}(0,0)=0$.} and $a'(0)=\frac{da}{dy}(0)$. The approximate expression of $I(y_0,y)$ is therefore given at order 2 by

\begin{align}
I(y_0,y)&=\int_{y_0}^{y}e^{-\frac{2}{\epsilon}\left(A(0)+\frac{y^2}{2}a'(0)\right)}dy\nonumber\\
&=\frac{1}{2}\sqrt{\frac{\pi \epsilon}{a'(0)}}e^{-\frac{2}{\epsilon}A(0)}\nonumber\\
&\times\left(\erf\left(y\sqrt{\frac{a'(0)}{\epsilon}}\right)-\erf\left(y_0\sqrt{\frac{a'(0)}{\epsilon}}\right)\right)\label{eq:Iappb}
\end{align}

where $\erf$ is the error function.

Since we consider $0<\epsilon\ll 1$ and because $a'(0)$ has a finite value, one has for an initial value chosen smaller than the static bifurcation point, i.e. for $y_0<0$
\begin{equation}
y_0\sqrt{\frac{a'(0)}{\epsilon}}\ll-1 \quad \Rightarrow \quad \erf\left(y_0\sqrt{\frac{a'(0)}{\epsilon}}\right) \approx -1
\label{eq:assum1}
\end{equation}
and for $y>0$ (we are interested in the dynamic bifurcation point which is by definition larger than the static bifurcation point) 
\begin{equation}
y\sqrt{\frac{a'(0)}{\epsilon}}\gg1 \quad \Rightarrow \quad \erf\left(y\sqrt{\frac{a'(0)}{\epsilon}}\right) \approx 1.
\label{eq:assum2}
\end{equation}

Therefore, from Eqs.~\eqref{eq:assum1} and \eqref{eq:assum2}, Eq.~\eqref{eq:Iappb} reduces to 
\begin{equation}
I(y_0,y)=\sqrt{\frac{\pi \epsilon}{a'(0)}}e^{-\frac{2}{\epsilon}A(0)}.
\label{eq:ExpI1}
\end{equation}

Eq.~\eqref{eq:ExpI1} highlights that $I(y_0,y)$ is now independent of $y$ and $y_0$ and simply denoted $I$. Therefore Eq.~\eqref{eq:eqDynPtStoch1} becomes
\begin{equation}
A(y)=K,
\label{eq:eqDynPtStoch2}
\end{equation}
with 
\begin{equation}
K=-\epsilon\ln\sigma-\frac{\epsilon}{2}\ln\frac{I}{\epsilon}+\epsilon\ln x_{y_0}
\end{equation}
and the function $A(y)$ given by \eqref{eq:defA}. 

Eq.~\eqref{eq:eqDynPtStoch2} can be expressed as a third order polynomial equation with respect to $y$ as
\begin{equation}
a_1y^3+a_2y^2+a_3y+a_4=0
\label{eq:cubic}
\end{equation}
with
\begin{equation}
\begin{split}
a_1&=\frac{\zeta ^2 F_1^2}{4 \omega _1^2},\\
a_2&=\frac{1}{4} \left(\frac{(3 \hat{\gamma}^{\text{st}} -2) \zeta ^2 F_1^2}{\omega _1^2}-\alpha _1^2\right),\\
a_3&=\frac{(\hat{\gamma}^{\text{st}} -1) (3 \hat{\gamma}^{\text{st}} -1) \zeta ^2 F_1^2}{4 \omega _1^2}-\alpha _1 K,\\
a_4&=\frac{(\hat{\gamma}^{\text{st}} -1)^2 \hat{\gamma}^{\text{st}}  \zeta ^2 F_1^2}{4 \omega _1^2}-K^2
\end{split}
\end{equation}
and $\hat{\gamma}^{\text{st}}$ given by Eq.~\eqref{eq:statbifpt} (details are given in  Appendix~\ref{app:stochDyn}).

Eq.~\eqref{eq:cubic} is solved analytically using Cardano's formula (see again Appendix~\ref{app:stochDyn} for details). One obtains
\begin{subnumcases}{\label{eq:stocDBPa}\hat{y}^{\text{dyn}}_{\text{stoch,a}}=}
r_0 & if $\sigma<\sigma_3$, \\
r_2 & if $\sigma_3<\sigma<\sigma_2$ 
\end{subnumcases}
where $r_0$ and $r_2$ are given by Eq.~\eqref{eq:cardansol1}, as depicted in Fig.~\ref{fig:pSoluCard}. The expressions of $\sigma_2$ and $\sigma_3$ are given respectively by Eqs.~\eqref{eq:sigma1} to \eqref{eq:sigma3}.

A second approximate expression of the stochastic dynamic bifurcation point is obtained using \eqref{eq:TSAb} in Eq.~\eqref{eq:eqDynPtStoch2} which becomes a second order polynomial equation with respect to $y$. Solving the latter leads to the following less accurate approximate, but easier to interpret, expression
\begin{align}
\hat{y}^{\text{dyn}}_{\text{stoch,b}}&=
2 \Bigg[
-\frac{(\hat{\gamma}^{\text{st}})^{3/2} \omega _1 \epsilon}{\zeta  F_1(3 \hat{\gamma}^{\text{st}} +1) }\nonumber\\
&\times\Bigg(\frac{3 \ln (\hat{\gamma}^{\text{st}})}{2}
+\ln \left(\frac{8 \pi  \omega _1}{\epsilon\zeta  F_1(3 \hat{\gamma}^{\text{st}} +1) }\right)\nonumber\\
&\quad+4 \ln (\sigma )-4 \ln \left(x_{y_0}\right)\Bigg)\Bigg]^{1/2}.
\label{eq:stocDBPb}
\end{align}

Note that both in Eqs.~\eqref{eq:stocDBPa} and \eqref{eq:stocDBPb} the dependence on the initial value $y_0$ is lost. However, contrary to the deterministic case, the stochastic dynamic bifurcation point depends on the slope $\epsilon$.

\begin{figure}[t!]
	\centering
	\includegraphics[width=1\columnwidth]{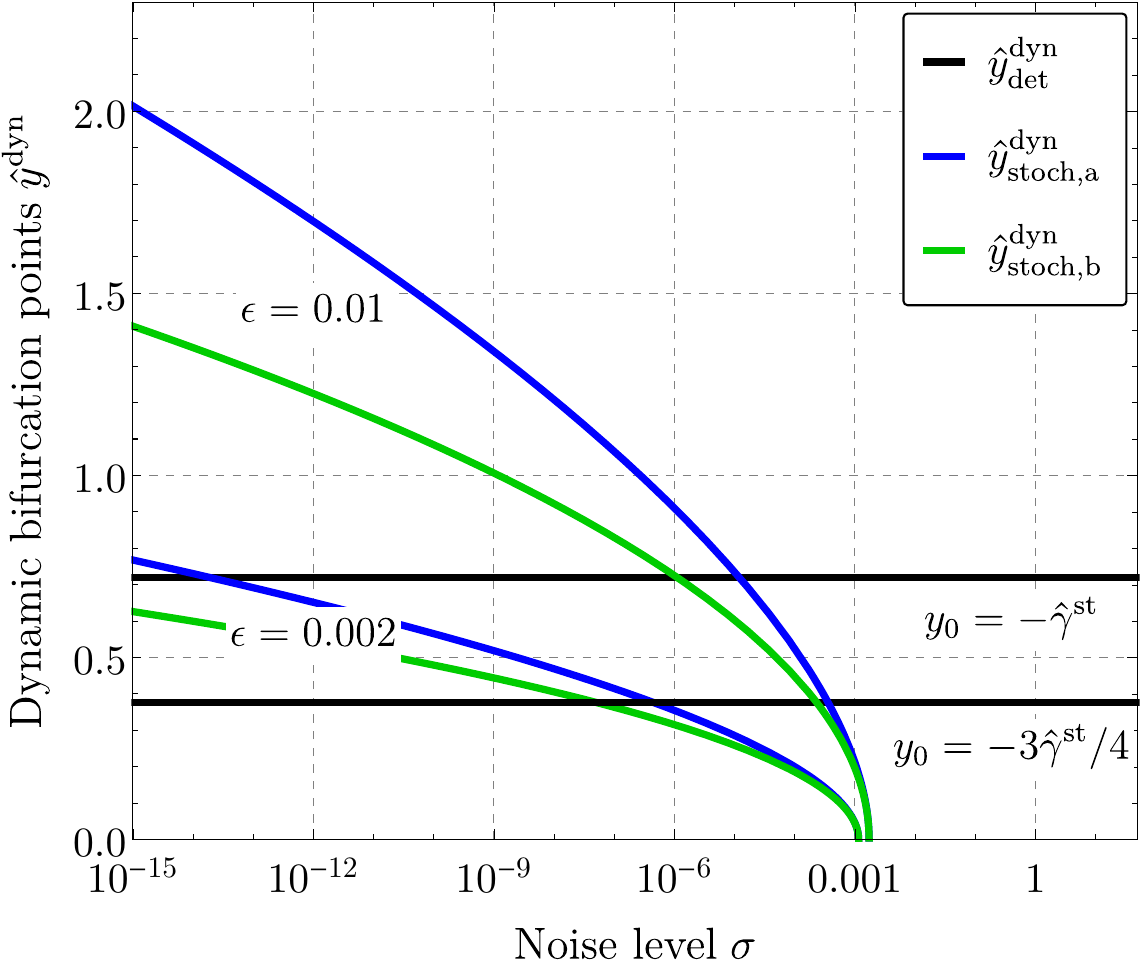}
	\caption{Deterministic and stochastic dynamic bifurcation points, given by Eqs.~\eqref{eq:dyndetbifpt1}, \eqref{eq:stocDBPa} and \eqref{eq:stocDBPb} respectively, as functions of the noise level $\sigma$. The deterministic bifurcation point $\hat{y}^{\text{dyn}}_{\text{det}}$ is plotted for two values of the initial condition $y_0$, i.e. $y_0=-\hat{\gamma}^\text{st}$ and $-3\hat{\gamma}^\text{st}/4$. The stochastic bifurcation points are plotted for two values of the parameter $\epsilon$, i.e. $\epsilon=0.002$ and $0.01$. The other parameters are given by \eqref{eq:param1} with in addition $x_{y_0}=0.01$.}
	\label{fig:compbifdyn}
	%param = {\[Alpha]1 -> 0.02`100, \[Omega]1 -> 1000`100,   F1 -> 1200`100, \[Epsilon] -> 0.002`100, \[Zeta] -> 0.2`100};
\end{figure}

Deterministic and stochastic dynamic bifurcation points, given by Eqs.~\eqref{eq:dyndetbifpt1}, \eqref{eq:stocDBPa} and \eqref{eq:stocDBPb} respectively, are plotted in Fig.~\ref{fig:compbifdyn} as functions of the noise level $\sigma$. The deterministic bifurcation point $\hat{y}^{\text{dyn}}_{\text{det}}$ is plotted for two values of the initial condition $y_0$, i.e. $y_0=-\hat{\gamma}^\text{st}$ and $-3\hat{\gamma}^\text{st}/4$. The stochastic bifurcation points are plotted for two values of the parameter $\epsilon$, i.e. $\epsilon=0.002$ and $0.01$. The other parameters are given by \eqref{eq:param1} with in addition $x_{y_0}=0.01$. First, we can see that the higher the noise level the closer the two approximate expressions of the stochastic dynamic bifurcation points. This is because the higher the noise level the more the bifurcation delay is reduced, hence the more the Taylor series \eqref{eq:TSAb} is valid. Secondly, Fig.~\ref{fig:compbifdyn} shows graphically the domains of existence of the regimes listed at the end of the Sect.~\ref{sec:EVA}.

The following definitions are chosen for the boundary values (with respect to the noise level $\sigma$) between the different regimes previously mentioned: the boundary value between Regime I and Regime II, denoted $\sigma_{I/II}$, corresponds to the intersection between $\hat{y}^{\text{dyn}}_{\text{det}}$ and $\hat{y}^{\text{dyn}}_{\text{stoch,a}}$. This means in Fig.~\ref{fig:compbifdyn} that if $\sigma<\sigma_{I/II}$ (Regime I), $\hat{y}^{\text{dyn}}_{\text{det}}$ is the actual dynamic bifurcation point. On the contrary, if $\sigma>\sigma_{I/II}$ (Regime II),  $\hat{y}^{\text{dyn}}_{\text{stoch,a}}$ or  $\hat{y}^{\text{dyn}}_{\text{stoch,b}}$ (depending on the approximation retained) is the actual dynamic bifurcation point. The boundary value between Regime II and Regime III, denoted $\sigma_{II/III}$, is the value at which $\hat{y}^{\text{dyn}}_{\text{stoch,a}}$ (and $\hat{y}^{\text{dyn}}_{\text{stoch,b}}$) no longer exists. The expressions of $\sigma_{I/II}$ and $\sigma_{II/III}$ are given in next section.

 %(resp. $\sigma>\sigma_{I/II}$) 
 
  %(resp.  $\hat{y}^{\text{dyn}}_{\text{stoch,a}}$ or  $\hat{y}^{\text{dyn}}_{\text{stoch,b}}$, it depends on the approximation we retained)

%_____________________________Subsection_____________________________%
\subsection{Domains of existence of the regimes}\label{sec:opreg}

The boundary value between Regime I and Regime II, with respect to the noise level $\sigma$ and denoted $\sigma_{I/II}$, is obtained solving
\begin{equation}
\mathcal{D}(y) =\mathcal{S}(y).
\end{equation}
with respect to $\sigma$. Therefore, from \eqref{eq:EVSA2b} and \eqref{eq:ExpI1}, we find
\begin{equation}
\sigma_{I/II}=x_{y_0}\left(\frac{a'(0)\epsilon}{\pi}\right)^{1/4}
e^{\frac{2}{\epsilon}\left(A(0)-A(y_0)\right)}.
\label{eq:siqI-II}
\end{equation}

\begin{figure*}[t!]
	\centering
	\includegraphics[width=1.9\columnwidth]{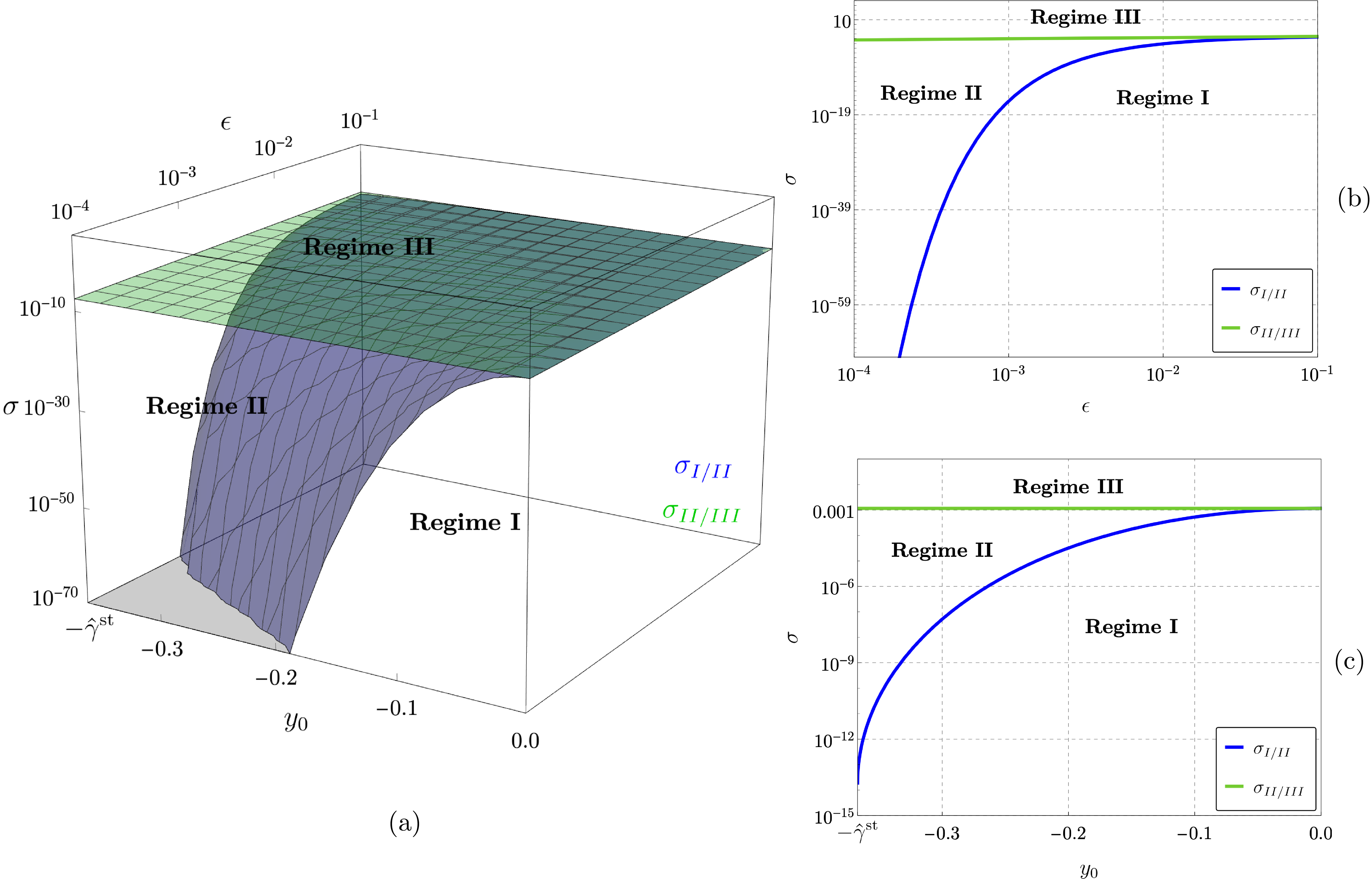}
	\caption{Regions of existence of the Regimes I, II and III (a) in the $(y_0,\epsilon,\sigma)$-space, (b) in the $(y_0,\sigma)$-plane for $\epsilon=0.002$ and (c) in the plane $(\epsilon,\sigma)$-plane for $y_0=-0.34$. The expressions of $\sigma_{I/II}$ and $\sigma_{II/III}$ are given by Eqs.~\eqref{eq:siqI-II} and \eqref{eq:siqII-III} respectively.}
	\label{fig:regimes1}
	%param = {\[Alpha]1 -> 0.02`100, \[Omega]1 -> 1000`100,   F1 -> 1200`100, \[Epsilon] -> 0.002`100, \[Zeta] -> 0.2`100};
\end{figure*}

The second boundary value between Regime II and Regime III, again with respect to the noise level $\sigma$ denoted $\sigma_{II/III}$, corresponds to the value of $\sigma$ for which $\hat{y}^{\text{dyn}}_{\text{stoch,a}}$ and $\hat{y}^{\text{dyn}}_{\text{stoch,b}}$ do not exist anymore\footnote{This corresponds to the fact that the solutions of Eq.~\eqref{eq:eqDynPtStoch2} for both cubic and quadratic approximations become complex for $\sigma>\sigma_{II/III}$.} (see Fig.~\ref{fig:compbifdyn}). The expression of $\sigma_{II/III}$ is obtained by noting that $\hat{y}^{\text{dyn}}_{\text{stoch,a}}$ and $\hat{y}^{\text{dyn}}_{\text{stoch,b}}$ vanish at $\sigma=\sigma_{II/III}$. Therefore, one may solve $\hat{y}^{\text{dyn}}_{\text{stoch,b}}=0$ that, from \eqref{eq:stocDBPb}, leads to
\begin{equation}
\sigma_{II/III}=
x_0\left(\frac{1}{\hat{\gamma}^{\text{st}}}\right)^{3/8}\left(\frac{\epsilon\zeta  F_1(3 \hat{\gamma}^{\text{st}} +1) }{8 \pi\omega _1}\right)^{1/4}
\label{eq:siqII-III}
\end{equation}
with $\hat{\gamma}^{\text{st}}$ given by Eq.~\eqref{eq:statbifpt}.

Using Eqs.~\eqref{eq:siqI-II} and \eqref{eq:siqII-III} respectively the values of $\sigma_{I/II}$ and $\sigma_{II/III}$ corresponding to situations plotted in Fig.~\ref{fig:compbifdyn} are given in Tab.~\ref{tab:sigRed}.

\begin{table}[t!]
\centering
\caption{Values $\sigma_{I/II}$ and $\sigma_{II/III}$, computed through Eqs.~\eqref{eq:siqI-II} and \eqref{eq:siqII-III} respectively, corresponding to situations plotted in Fig.~\ref{fig:compbifdyn}.}
\label{tab:sigRed}
\begin{tabular}{c|l|cc}
\hline
&&$y_0=-\hat{\gamma}^\text{st}$&$y_0=-3\hat{\gamma}^\text{st}/4$\\
\hline
\multirow{2}{*}{$\sigma_{I/II}$} & $\epsilon=0.002$ &  $1.88\cdot10^{-14}$ &  $4.39\cdot10^{-7}$\\
& $\epsilon=0.01$ & $1.20\cdot10^{-5}$ & $3.59\cdot10^{-4}$\\\hline
\multirow{2}{*}{$\sigma_{II/III}$} & $\epsilon=0.002$  & $1.16\cdot10^{-3}$ &   $1.16\cdot10^{-3}$\\
 & $\epsilon=0.01$ & $1.17\cdot10^{-3}$ &   $1.17\cdot10^{-3}$\\\hline
\end{tabular}
\end{table}

By means of Eqs.~\eqref{eq:siqI-II} and \eqref{eq:siqII-III} one can also depict the regions of existence of each of the three regimes in the $(y_0,\epsilon,\sigma)$-space (see Fig.~\ref{fig:regimes1}(a)). A representation of these regions is also given in the $(y_0,\sigma)$-plane for $\epsilon=0.002$ and in the plane $(\epsilon,\sigma)$-plane for $y_0=-0.34$ in Figs.~\ref{fig:regimes1}(b) and \ref{fig:regimes1}(c) respectively. One can see that the boundary between Regime II and Regime III mainly depends on $\sigma$. On the contrary, the boundary between Regime I and Regime II depends simultaneously on $\sigma$, $y_0$ and $\epsilon$. That means that the use of Eq.~\eqref{eq:dyndetbifpt1} (the deterministic bifurcation point) or Eq.~\eqref{eq:stocDBPb} (the stochastic bifurcation point) as a theoretical prediction of the actual dynamic bifurcation point observed on numerical simulations does not depend only on the noise level. Fig.~\ref{fig:regimes1} shows that the dependence on $\epsilon$ is the most important. Indeed, the above value $\sigma_{I/II}=3.15\cdot10^{-10}$ has been obtained for $\epsilon=0.002$. The value becomes $\sigma_{I/II}=7.21\cdot10^{-17}$ for $\epsilon=10^{-3}$ and $\sigma_{I/II}=1.44\cdot10^{-69}$ for $\epsilon=2\cdot 10^{-4}$. In other words, if $\epsilon=2\cdot 10^{-4}$, to simulate the deterministic case, the number of digits of precision used by the computer must be at least equal to 69.

The latter observation shows that, in some instances, even with an extremely low level, the noise can influence the bifurcation delay. That means for example that a deterministic approach can fail to predict the bifurcation delay observed on numerical integration of the model in which the round-off errors act as an additive noise with a very low level~\cite{BergeotNLD2012b}.

%_____________________________Subsection_____________________________%
\subsection{Probability density function of the stochastic averaged amplitude}

In addition to being able to obtain the expected value of the squared amplitude~\eqref{eq:EVSA2}, Eq.~\eqref{eq:solPitchGeby2} allows us to calculate the probability density function (PDF) of the amplitude $x_y$ without having to solve the Fokker-Planck equation associated to Eq. \eqref{eq:VdP1Moy3Stochay}. Indeed, a known result of stochastic calculus (see e.g. \cite{bookFima2005}, Chap. 4)  is that  the Itô integral of a deterministic function $f(t)$, i.e. $\int_0^tf(t)dW_t$,
is a Gaussian process with mean equal to zero and variance $v=\int_0^tf(t)^2dt$. Therefore, from \eqref{eq:solPitchGeby2}, $x_y$ is a Gaussian process with mean $\EE{x_{y}}=x^\text{det}(y)$ and variance
\begin{equation}
\EE{x_{y}^2}-\EE{x_{y}}^2=
\mathcal{S}(y)=\frac{\sigma^2}{\epsilon}e^{\frac{2}{\epsilon}A(y)}\int_{y_0}^{y}e^{-\frac{2}{\epsilon}A(y)}dy
\end{equation}
with $\int_{y_0}^{y}e^{-\frac{2}{\epsilon}A(y)}dy$ given by \eqref{eq:ExpI1}.

The associated Gaussian PDF is therefore
\begin{equation}
\rho(x,y)=\frac{1}{\sqrt{2 \pi\mathcal{S}(y)}}e^{-\frac{(x-x^\text{det}(y))^2}{2 \mathcal{S}(y)}}.
\label{eq:PDE2}
\end{equation}
%In the Regime II we have $\mathcal{D}(y)\ll\mathcal{S}(y)$ and \eqref{eq:PDE1} becomes
%\begin{equation}
%\rho(x,y)=\frac{1}{\sqrt{2 \pi} \mathcal{S}(y)}e^{-\frac{x^2}{2 \mathcal{S}(y)^2}}.
%\label{eq:PDE2}
%\end{equation}
This means that, for a given value of $y$, the probability that $x_y$ lies between $x$ and $x+dx$ is $\rho(x,y)dx$. Then, we define the function $R(y)=\rho(x_{y_0},y)$ for which the probability that $x_y$ lies between $x_{y_0}$ and $x_{y_0}+dx$ is $R(y)dx$. From \eqref{eq:PDE2}, the expression of the function $R(y)$ is
\begin{equation}
R(y)=\frac{1}{\sqrt{2 \pi \mathcal{S}(y)}}e^{-\frac{(x_{y_0}-x^\text{det}(y))^2}{2 \mathcal{S}(y)}}.
\label{eq:PDE4}
\end{equation}

In the Regime I we have $\mathcal{D}(y)=x^\text{det}(y)^2\gg\mathcal{S}(y)$ and the function $R(y)$ can be approximated by a Dirac delta function as $R(y)\approx \delta(x_{y_0}-x^\text{det}(y))$. Denoting $\ell(y)=x_{y_0}-x^\text{det}(y)$ the roots of which are $y_0$ and $\hat{y}^{\text{dyn}}_{\text{det}}$ (see Eq.~\eqref{eq:AA0}) and using the composition property of Dirac delta function one obtains
\begin{equation}
R(y)\approx R_I(y)
%=\delta(\ell(y))
=\frac{\delta(y-y_0)}{|\ell'(y_0)|}+\frac{\delta(y-\hat{y}^{\text{dyn}}_{\text{det}})}{|\ell'(\hat{y}^{\text{dyn}}_{\text{det}})|}
\label{eq:PDE5}
\end{equation}
when the system evolves according to Regime I.

In the Regime II, because $\mathcal{D}(y)\ll\mathcal{S}(y)$, Eq.~\eqref{eq:PDE4} becomes
\begin{equation}
R(y)\approx R_{II}(y)=\frac{1}{\sqrt{2 \pi \mathcal{S}(y)} }e^{-\frac{x_{y_0}^2}{2 \mathcal{S}(y)}}.
\label{eq:PDE6}
\end{equation}
We have therefore
\begin{equation}
R_{II}'(y)=\frac{\mathcal{S}'(y)\left(x_{y_0}^2- \mathcal{S}(y)\right)}{\sqrt{2 \pi} \mathcal{S}(y)^{5/2}}e^{-\frac{x_{y_0}^2}{2 \mathcal{S}(y)}}.
\label{eq:PDE7}
\end{equation}
which vanishes for $x_{y_0}= \mathcal{S}(y)$ (it can be shown that this corresponds to a maximum of $R_{II}'(y)$) and therefore, according to the Definition~\ref{def:ptbifdyn}, for $y=\hat{y}^{\text{dyn}}_{\text{stoch,a}}$.

\begin{figure}[t!]
	\centering
	\subfigure[]{\includegraphics[width=1\columnwidth]{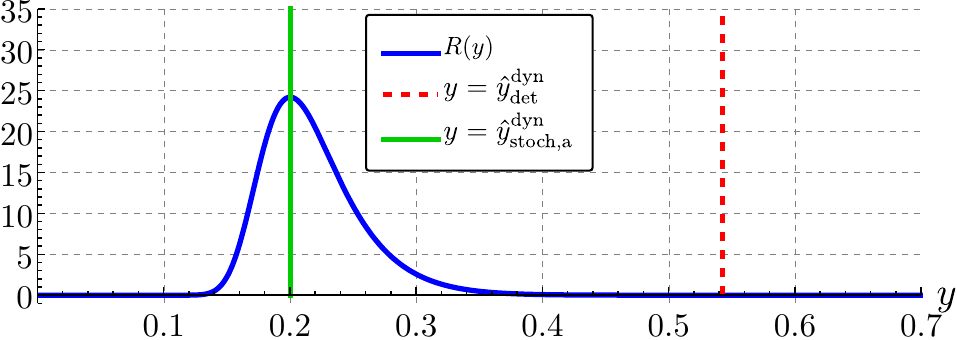}\label{fig:PDFa}}
	\subfigure[]{\includegraphics[width=1\columnwidth]{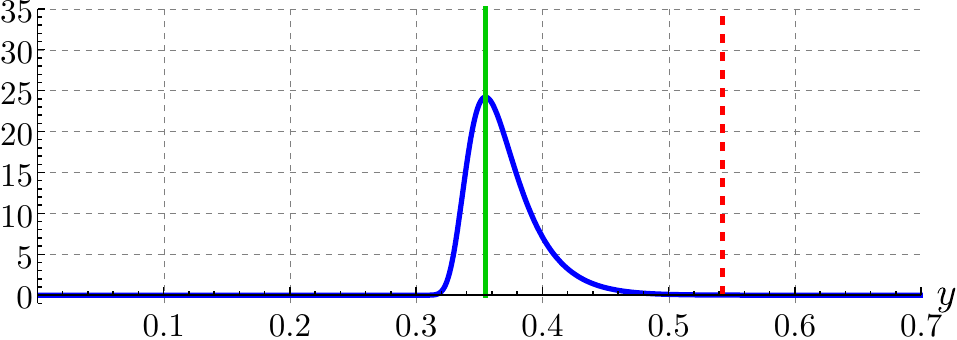}\label{fig:PDFb}}
	\subfigure[]{\includegraphics[width=1\columnwidth]{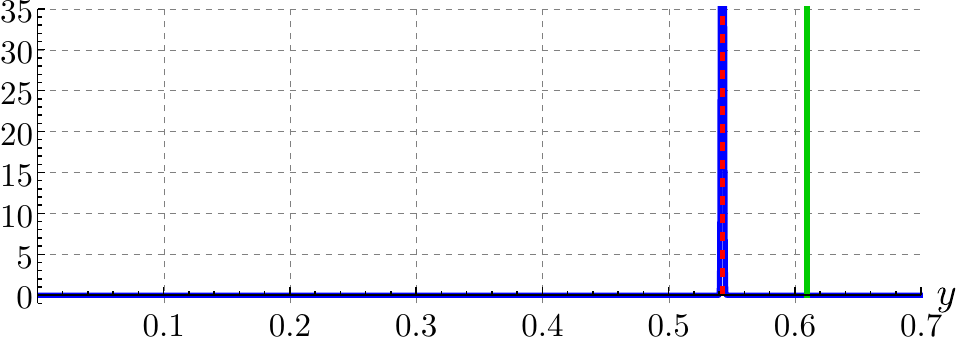}\label{fig:PDFc}}
	\caption{The function $R(y)$ defined by Eq.~\eqref{eq:PDE4}. The quantities $\hat{y}^{\text{dyn}}_{\text{det}}$, $\hat{y}^{\text{dyn}}_{\text{stoch,a}}$ and $\hat{y}^{\text{dyn}}_{\text{stoch,b}}$ are depicted by vertical green line and vertical red dashes line respectively. The parameters~\eqref{eq:param1} are used and (a) $\sigma=10^{-4}$, (b) $\sigma=10^{-6}$ and (c) $\sigma=10^{-11}$. Moreover $y_0=-0.34$, $x_{y_0}=0.01$ and $\epsilon=0.002$.}
\label{fig:PDF}
\end{figure}

In Fig.~\ref{fig:PDF} the function $R(y)$ is plotted using parameters~\eqref{eq:param1} and three values of the noise level, i.e. $\sigma=10^{-4}$, $\sigma=10^{-6}$ and $\sigma=10^{-11}$. For the first two values of the noise level the system is in Regime II and it is in the Regime I for the third value (see Fig. \ref{fig:pPosGen} in the next section). In Figs.~\ref{fig:PDFa} and \ref{fig:PDFb} one can see that the maximum of $R(y)$ is obtained for $y=\hat{y}^{\text{dyn}}_{\text{stoch,a}}$ as predicted by Eq.~\eqref{eq:PDE7}. Eq.~\eqref{eq:PDE5} is illustrated in Fig.~\ref{fig:PDFc} in which we can see that in Regime I the function $R(y)$ becomes a Dirac delta function translated to $y=\hat{y}^{\text{dyn}}_{\text{det}}$. In the figure only the positive values of $y$ are represented and $y_0<0$. Consequently, the first term of the sum in the right-hand side of Eq.~\eqref{eq:PDE5} does not appear.

%-------------------------------------------------------------------------------------------------%
%-------------------------------------------------------------------------------------------------%
% Section
%-------------------------------------------------------------------------------------------------%
%-------------------------------------------------------------------------------------------------%
\section{Comparison between theoretical results and numerical simulation}\label{Numerical_results}

%_____________________________Subsection_____________________________%
\subsection{Comparison in term of the expected value of the squared amplitude}

The aim of this section is to validate both the stochastic averaging procedure in term of expected value of the squared amplitude and the analytical predictions $\hat{y}^{\text{dyn}}_{\text{det}}$ of the deterministic dynamic bifurcation point given by Eq.~\eqref{eq:dyndetbifpt1} and those of the stochastic bifurcation points, $\hat{y}^{\text{dyn}}_{\text{stoch,a}}$ and $\hat{y}^{\text{dyn}}_{\text{stoch,b}}$, given by Eqs.~\eqref{eq:stocDBPa} and \eqref{eq:stocDBPb} respectively.

To achieve that, after getting the time series of $p_t$ from the numerical integration of~\eqref{eq:modclarstoch2}\footnote{All numerical simulations of the Itô stochastic differential equations are performed using the function \texttt{ItoProcess} of the \textit{Wolfram Mathematica} software.}, the corresponding amplitude (denoted $A_y$) is computed as a function of $y$. To that end, the time series of $p_t$ and $y_t$ are divided into $N$ intervals and on each of them the maximum of $p_t$ (max$(p)_i$ with $i=1,\dots,N$) and the mean of $y_t$ (mean$(y)_i$ with $i=1,\dots,N$) are computed. Then $A_y$ is defined as max$(p)_i$ as a function of mean$(y)_i$ for $i$ from 1 to $N$. The procedure is repeated 50 times and the expected value $\EE{A_y^2}$ is computed over the 50 realizations. Simultaneously, the expected value $\EE{x_y^2}$ of $x_y$ is computed over 50 realizations of the numerical integration of the averaged equation~\eqref{eq:VdP1Moy3Stochay}.

\begin{figure*}[t!]
	\centering
	\subfigure[]{\includegraphics[width=2\columnwidth]{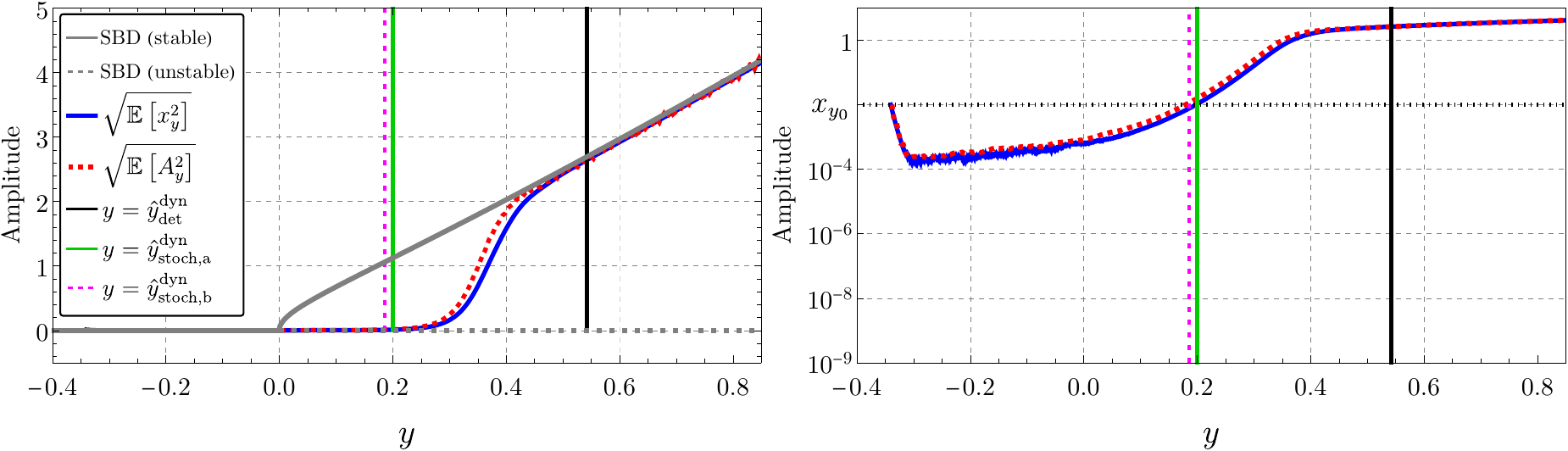}\label{fig:pDynBifPtvsSigVdPa}}
	\subfigure[]{\includegraphics[width=2\columnwidth]{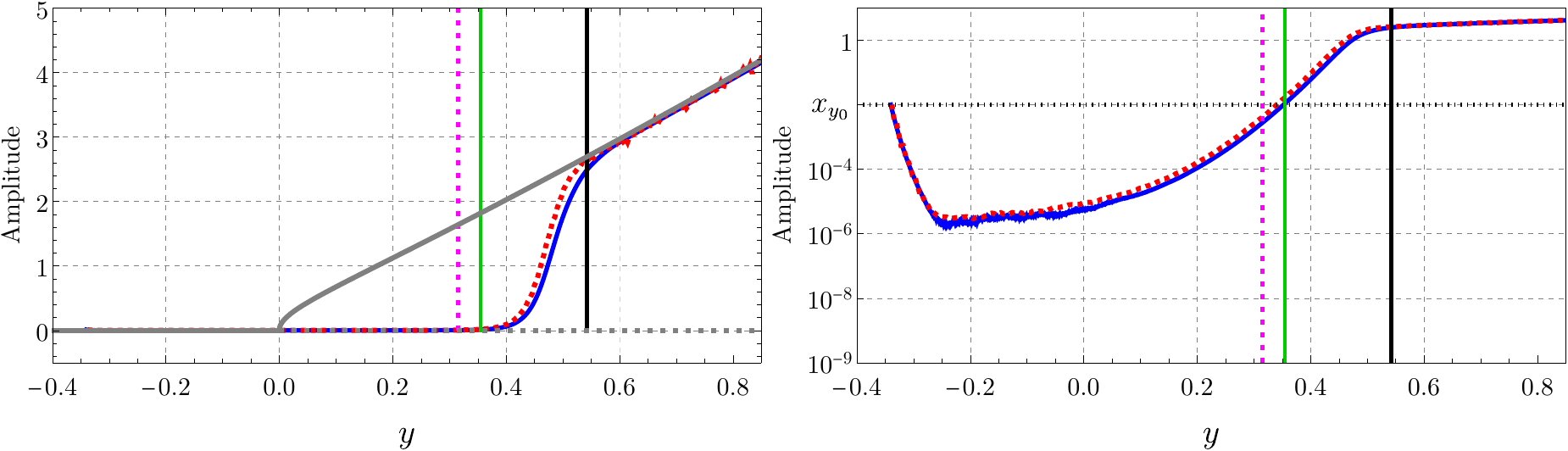}\label{fig:pDynBifPtvsSigVdPb}}
	\subfigure[]{\includegraphics[width=2\columnwidth]{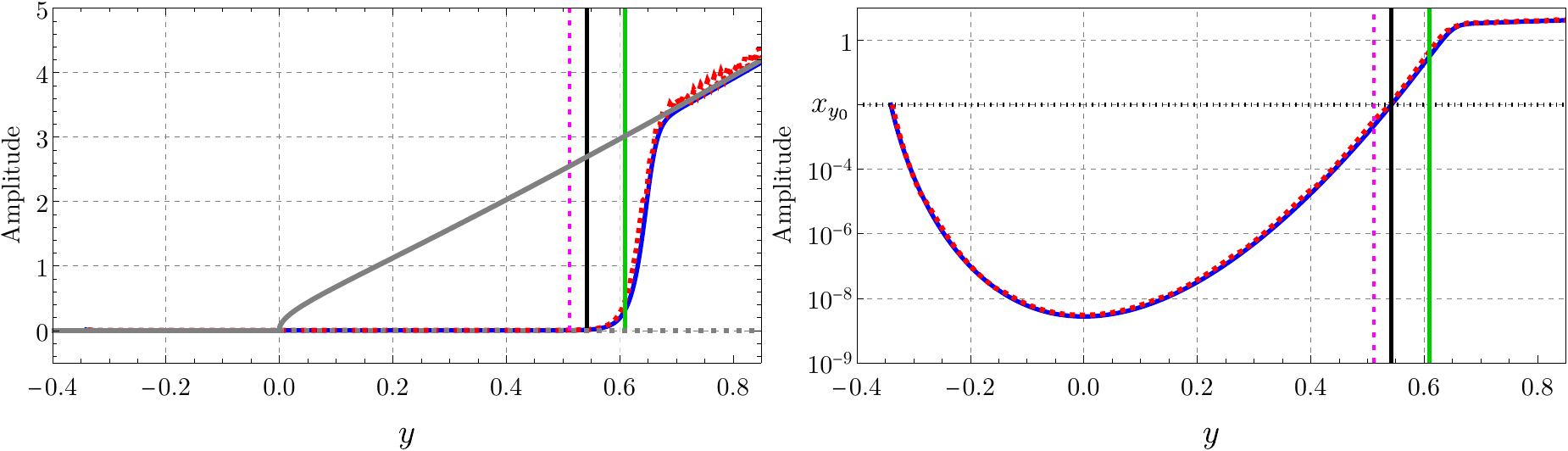}\label{fig:pDynBifPtvsSigVdPc}}
	\caption{On the left: a plot with the expected values of  $A_y^2$ (red dashed line) and $x_y^2$ (blue line) obtained using 50 realizations of the numerical integration of Eqs \eqref{eq:modclarstoch2} and \eqref{eq:VdP1Moy3Stochay} respectively, the stable branches of the static bifurcation diagram (SBD, gray line) computed in Appendix~\ref{app:bifdiag} and $\hat{y}^{\text{dyn}}_{\text{det}}$, $\hat{y}^{\text{dyn}}_{\text{stoch,a}}$ and $\hat{y}^{\text{dyn}}_{\text{stoch,b}}$ depicted by vertical black line, vertical green line and vertical magenta dashed line respectively. On the right: same as on the left without the static bifurcation diagram, with an horizontal black dashed at $\sqrt{\EE{x_y^2}}=x_{y_0}$ and using a logarithm scale for the $\sqrt{\EE{x_y^2}}$-axis. The set of parameters \eqref{eq:param1} is used and, from (a) to (c), $\sigma=10^{-4}$, $\sigma=10^{-6}$ and $\sigma=10^{-11}$. Moreover $y_0=-0.34$, $x_{y_0}=0.01$ and $\epsilon=0.002$.}
\label{fig:pDynBifPtvsSigVdP}
\end{figure*}

The results are shown in Fig.~\ref{fig:pDynBifPtvsSigVdP} which represents: (on the left) a plot of $\sqrt{\EE{A_y^2}}$ (dashed red line), $\sqrt{\EE{x_y^2}}$ (blue line), the static bifurcation diagram (red) computed in Appendix~\ref{app:bifdiag} and $\hat{y}^{\text{dyn}}_{\text{det}}$, $\hat{y}^{\text{dyn}}_{\text{stoch,a}}$ and $\hat{y}^{\text{dyn}}_{\text{stoch,b}}$ depicted by vertical black line, vertical green line and vertical magenta dashed line respectively; (on the right) same as on the left without the static bifurcation diagram, with an horizontal black dashed at $\sqrt{\EE{x_y^2}}=x_{y_0}$ and using a logarithm scale for the $\sqrt{\EE{x_y^2}}$-axis. The set of parameters \eqref{eq:param1} is used and, from top to bottom: $\sigma=10^{-4}$, $\sigma=10^{-6}$ and $\sigma=10^{-11}$. The initial conditions are $y_0=-0.34$ and $x_{y_0}=0.01$.

The agreement through numerical integration between expected of the squared amplitude of the stochastic slow dynamics and the initial full order system for different situations corresponding to Regimes I and II is considered as a successful assessment of the stochastic averaging procedure.

In Fig.~\ref{fig:pDynBifPtvsSigVdPa} and \ref{fig:pDynBifPtvsSigVdPb} the system is in Regime II. Therefore, the noise cannot be ignored and the dynamic bifurcation must be determined by $\hat{y}^{\text{dyn}}_{\text{stoch,a}}$ or $\hat{y}^{\text{dyn}}_{\text{stoch,b}}$. The figure shows that the latter provides a good approximation of the dynamic bifurcation point. Indeed, through the Definition~\ref{def:ptbifdyn}, the dynamic bifurcation point is the abscissa of the point of intersection between $\EE{x_y}$ and the horizontal at $\EE{x_y}=x_{y_0}$. The logarithmic scale used in  Fig.~\ref{fig:pDynBifPtvsSigVdP} (right column) allows us to locate this intersection and to see that the vertical green line, corresponding to $\hat{y}^{\text{dyn}}_{\text{stoch,a}}$, passes also through this intersection and the vertical magenta dashed line, corresponding to $\hat{y}^{\text{dyn}}_{\text{stoch,b}}$, passes a little on the left (all the more as $\sigma$ is decreased). Of course the expression $\hat{y}^{\text{dyn}}_{\text{stoch,a}}$ is more accurate than $\hat{y}^{\text{dyn}}_{\text{stoch,b}}$ because it corresponds to a lesser degree of approximation. In Fig.~\ref{fig:pDynBifPtvsSigVdPc} the system is in Regime I, consequently, the noise can be neglected and the dynamic bifurcation point is determined by $\hat{y}^{\text{dyn}}_{\text{det}}$. A good agreement between theoretical prediction and numerical simulations can still be observed on the right column. Indeed, the vertical black line passes through the point of intersection between $\EE{x_y}$ and the horizontal line at $\EE{x_y}=x_{y_0}$.

\begin{figure}[t!]
	\centering
	\includegraphics[width=1\columnwidth]{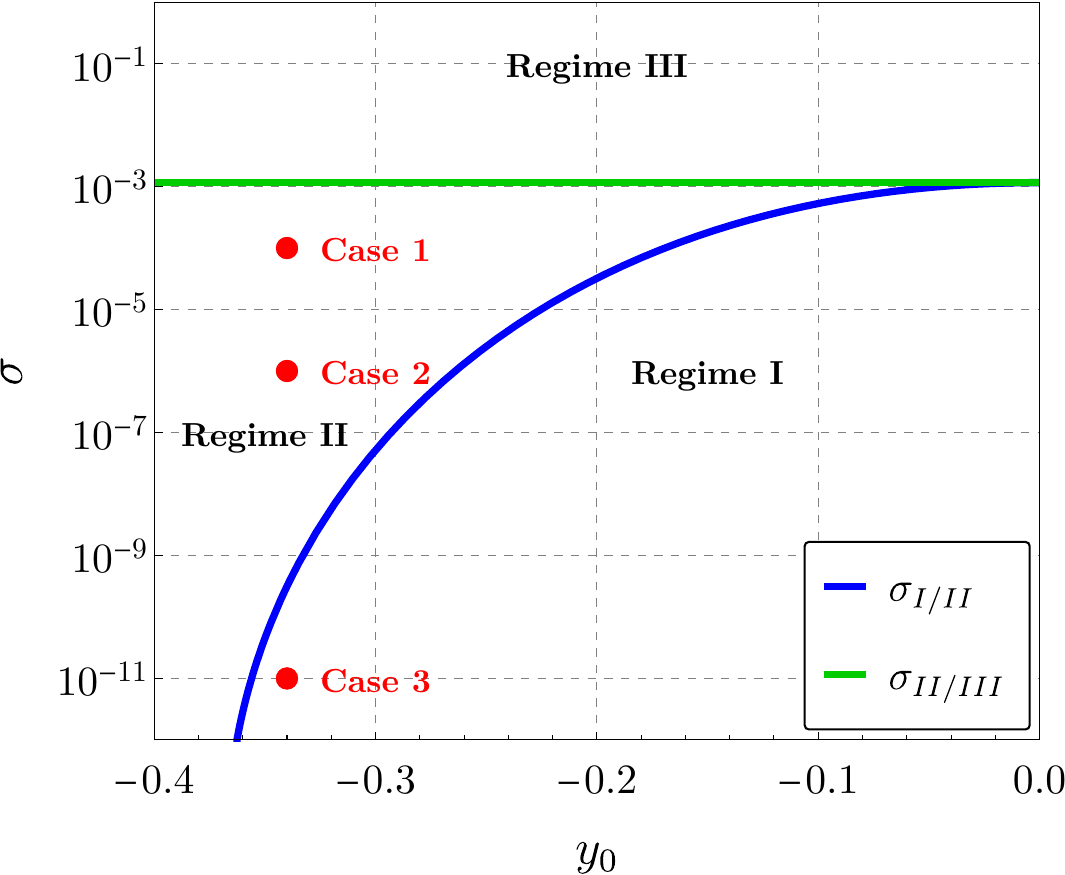}
	\caption{The regions of existence of the Regimes I, II and III in the $(y_0,\sigma)$-plane in which the red points show the position of the three cases depicted in Fig.~\ref{fig:pDynBifPtvsSigVdP}. The initial conditions are $y_0=-0.34$ and $x_{y_0}=0.01$ and three values of the noise level are used: $\sigma=10^{-4}$, $\sigma=10^{-6}$ and $\sigma=10^{-11}$ which correspond to cases 1 to 3 respectively.}
\label{fig:pPosGen}
\end{figure}

The position in the $(y_0,\sigma)$-plane of the three situations depicted in Fig.~\ref{fig:pDynBifPtvsSigVdP} is shown in Fig.~\ref{fig:pPosGen} and superimposed to the regions of existence of the Regimes I, II and III.

%_____________________________Subsection_____________________________%
\subsection{Comparison in term of probability density function}

\begin{figure}[t!]
	\centering
	\subfigure[]{\includegraphics[width=1\columnwidth]{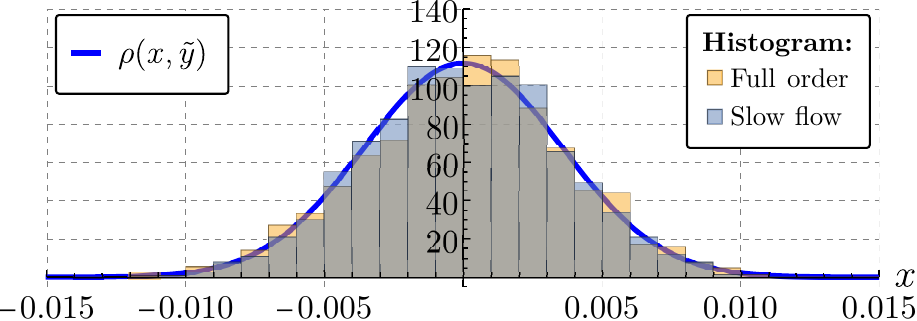}\label{fig:PDF2a}}
	\subfigure[]{\includegraphics[width=1\columnwidth]{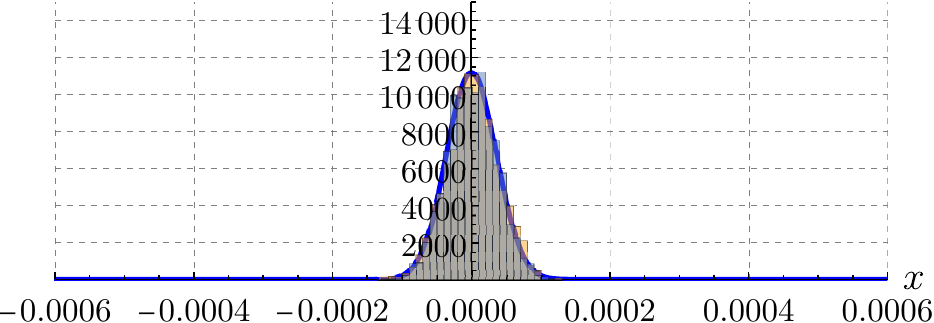}\label{fig:PDF2b}}
	\subfigure[]{\includegraphics[width=1\columnwidth]{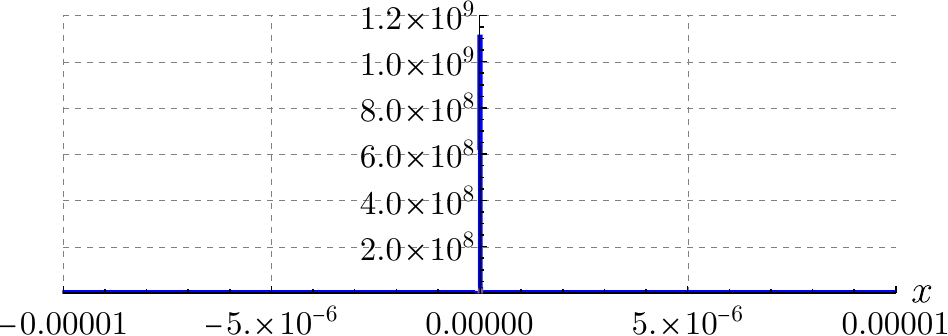}\label{fig:PDF2c}}
	\caption{Comparison between theoretical PDF $\rho(x,y)$ given by Eq.~\eqref{eq:PDE2} and two histograms: one in built from the full order system \eqref{eq:modclarstoch2} and the other from the stochastic slow dynamics~\eqref{eq:VdP1Moy3Stochay}. In both cases, 2000 realizations of the system are computed and we take, for each realization, the value of the considered random variable ($x_y$ or $p_t$) for a given value of $y$ denoted $\tilde{y}$ (chosen arbitrary to be $\tilde{y}=0.15$). The comparison is performed for three values of the noise level:  (a) $\sigma=10^{-4}$, (b) $\sigma=10^{-6}$ and (c) $\sigma=10^{-11}$. Moreover $y_0=-0.34$, $x_{y_0}=0.01$ and $\epsilon=0.002$.}
\label{fig:PDF2}
\end{figure}

In this section the theoretical PDF $\rho(x,y)$ (see Eq.\eqref{eq:PDE2}) is compared to two histograms: one in built from the full order system \eqref{eq:modclarstoch2} and the other from the stochastic slow dynamics~\eqref{eq:VdP1Moy3Stochay}. In both cases, 2000 realizations of the system are computed and we take, for each realization, the value of the considered random variable ($x_y$ or $p_t$) for a given value of $y$ denoted $\tilde{y}$ (chosen arbitrary to be $\tilde{y}=0.15$). The comparison, presented in Fig.~\ref{fig:PDF2}, is performed for three values of the noise level (the same as chosen previously in Figs.~\ref{fig:PDF} and \ref{fig:pDynBifPtvsSigVdP}) and shows en excellent agreement between theoretical and numerical results. In Fig.~\ref{fig:PDF2c}, the system is in Regime I and, as expected, the PDF appears as a Dirac delta function. In this case the histograms are also in agreement with the theoretical PDF but this cannot be seen in the figure.

We choose a value of $\tilde{y}$ smaller than the dynamic bifurcation point but results using a larger value (not presented here) shows also a good agreement between theoretical PDF and histograms as long as the linear approximation is valid and therefore for a value $y$ not too far from the dynamic bifurcation point.

%-------------------------------------------------------------------------------------------------%
%-------------------------------------------------------------------------------------------------%
% Section
%-------------------------------------------------------------------------------------------------%
%-------------------------------------------------------------------------------------------------%
\section{Discussion and conclusion}\label{sec:CCL}

This article is inspired by work initially done in the applied mathematics community.  The approach adopted is general and can be adapted to other systems undergoing a dynamic Hopf bifurcation.

For the clarinet model studied, the stochastic averaging procedure preserves the properties of the dynamic behavior of the original system. Thus, the same bifurcation delay is observed in the original and the averaged system. The different behaviors observed according to the amplitude of the random forcing, the initial condition or the rate of change of the bifurcation parameter are in line with those found in other fields of physics or in applied mathematics. In particular, the conclusions are the same as for another simple clarinet model already studied a few years ago. The latter model is a discrete-time model and is based on very different simplifying assumptions: it results from the discretization of a delay model which therefore retains an infinite number of acoustic modes. In contrast, the model studied in this article retains only a single acoustic mode. 

Studying the dynamic bifurcations of ODE-based models is nevertheless necessary. Firstly, because these are the models used by researchers in the context of instrument making. Secondly, because recent works underline the importance of taking into account the temporal dynamics of the bifurcation parameters (in particular those controlled by the musician~\cite{ColinotPHD2020}). The study carried out in this article already allows to obtain analytically the dynamic oscillation thresholds according to the main parameters controlled by the musician. This information could be linked to important notions for the musician and the instrument manufacturer such as the ease of playing. This opens interesting perspectives such as the study of dynamic bifurcations of a more realistic model than the one studied in this article. The first improvement is undoubtedly to complete this model with other acoustic modes whose importance is already known in the static case~\cite{ChaiKergoEn2016}.

%-------------------------------------------------------------------------------------------------%
%-------------------------------------------------------------------------------------------------%
% Section
%-------------------------------------------------------------------------------------------------%
%-------------------------------------------------------------------------------------------------%
\section*{Declarations}

\paragraph{\normalfont\bfseries Funding} Not applicable.

\paragraph{\normalfont\bfseries Conflict of interest}

The authors declare that they have no conflict of interest concerning the publication of this manuscript.

\paragraph{\normalfont\bfseries Availability of data and material} On request.

\paragraph{\normalfont\bfseries Code availability} On request.

%-------------------------------------------------------------------------------------------------%
%-------------------------------------------------------------------------------------------------%
%-------------------------------------------------------------------------------------------------%
%-------------------------------------------------------------------------------------------------%
% Appendices
%-------------------------------------------------------------------------------------------------%
%-------------------------------------------------------------------------------------------------%
%-------------------------------------------------------------------------------------------------%
%-------------------------------------------------------------------------------------------------%
\appendix
\normalsize

%-------------------------------------------------------------------------------------------------%
%-------------------------------------------------------------------------------------------------%
% Section
%-------------------------------------------------------------------------------------------------%
%-------------------------------------------------------------------------------------------------%
\section{The 1-dimensional Itô’s formula}\label{app:ItoFormula}

%See for example Theorem 4.1.2 in \cite{bernt2003}.

Let the following Itô differential equation
\begin{equation}
dx_t=m(x_t,t) dt+\sigma(x_t,t) dW_t,
\end{equation}
where $m$ and $\sigma$ are real functions and $W_t$ is the so-called Wiener process. 

Let $f(x_t,t)\in \mathcal{C}^{2}(\mathbb{R},\mathbb{R}^+)$ (i.e. $f$ is twice continuously differentiable on $(\mathbb{R},\mathbb{R}^+)$). Then $f(x_t,t)$ is also Itô process, and
\begin{subequations}
\label{eq:ItoFormula}
\begin{align}
df(x_t,t) &=\frac{\partial f}{\partial t}dt + \frac{\partial f}{\partial x}dx_t + \frac{1}{2}\frac{\partial^2 f}{\partial x^2}\sigma(x_t,t)^2dt\\
&= \left(\frac{\partial f}{\partial t} + m(x_t,t) \frac{\partial f}{\partial x} + \frac{\sigma(x_t,t) ^2}{2}\frac{\partial^2 f}{\partial x^2}\right)dt\nonumber\\
&+ \sigma(x_t,t) \frac{\partial f}{\partial x}\,dW_t.
\end{align}
\end{subequations}

Eq.~\eqref{eq:ItoFormula} is the 1-dimensional Itô’s formula (for more details and proof see~\cite{bernt2003}, Chap. 4).

%-------------------------------------------------------------------------------------------------%
%-------------------------------------------------------------------------------------------------%
% Section
%-------------------------------------------------------------------------------------------------%
%-------------------------------------------------------------------------------------------------%
\section{General formulation of the stochastic averaging method}\label{app:3}

In this appendix the stochastic averaging method~\cite{stratonovich1967topics,Khasminskii1966} is briefly described. For that we consider the following system of differential equations in standard form
\begin{equation}
\dot{\bf x}_t={\bf f}({\bf x}_t,t)+{\bf g}({\bf x}_t,t)\boldsymbol{\eta}_t
\label{eq:stochave1}
\end{equation}
where ${\bf x}_t \in \mathbb{R}^n$. If the deterministic vector function ${\bf f}({\bf x}_t,t)\in \mathbb{R}^n$ and matrix function ${\bf g}({\bf x}_t,t)\in \mathbb{R}^n \times  \mathbb{R}^n$ satisfy certain requirements~\cite{Khasminskii1966} and the elements of the vector $\boldsymbol{\eta}_t$ are broadband processes, with zero means, then the slow (or averaged) dynamics of~Eq.~\eqref{eq:stochave1} may be approximated by the following Itô equations
\begin{equation}
 d {\bf x}_t={\bf m}({\bf x}_t ) dt+\boldsymbol{\sigma}({\bf x}_t,t) d{\bf W}_t,
\label{eq:stochave2}
\end{equation}
where ${\bf W}_t\in \mathbb{R}^n$ is a vector of $n$ Wiener processes. The vector $\bf m$ and the matrix $\boldsymbol{\sigma}$ are called \textit{drift vector} and \textit{diffusion matrix} respectively and defined by
\begin{equation}
{\bf m} = T^{\text{av}}
\left\lbrace
{\bf f}+
\int_{-\infty}^0
\EE{
\left(\frac{\partial{(\bf g \boldsymbol{\eta})}}{\partial{\bf x} } \right)_t ({\bf g} \boldsymbol{\eta})_{t+\tau}}d\tau
\right\rbrace,
\label{eq:stochavem}
\end{equation}
where $\frac{\partial{(\bf g \boldsymbol{\eta})}}{\partial{\bf x}}$ is the Jacobian matrix of $\bf g \boldsymbol{\eta}$, and
\begin{equation}
\boldsymbol{\sigma} \boldsymbol{\sigma}^T=
 T^{\text{av}}
\left\lbrace
\int_{-\infty}^{+\infty}
\EE{
({\bf g} \boldsymbol{\eta})_{t}
({\bf g} \boldsymbol{\eta})_{t+\tau}^T}d\tau
\right\rbrace
\label{eq:stochavesig}
\end{equation}
where $\{.\}^T$ and $\EE{\{.\}}$ denotes respectively the transpose and the expected value of $\{.\}$. $ T^{\text{av}}$ is an averaging operator defined as follows
\begin{equation}
 T^{\text{av}}
\left\lbrace
.
\right\rbrace=
\lim\limits_{T \rightarrow +\infty} \frac{1}{T}\int_{t_0}^{t_0+T}\left\lbrace .\right\rbrace dt.
\label{eq:stochaveTav}
\end{equation}
It should be noted that in the case of a periodic variables with period $T_0$ (which is the case in this paper), the operator  $T^{\text{av}}$ becomes a classical Krylov–Bogolyubov time averaging over one period $T_0$, i.e.
\begin{equation}
 T^{\text{av}}
\left\lbrace
.
\right\rbrace=
\frac{1}{T_0}\int_{t_0}^{t_0+T_0}\left\lbrace .\right\rbrace dt
\label{eq:stochaveTav}
\end{equation}
and the result is independent of $t_0$.

%-------------------------------------------------------------------------------------------------%
%-------------------------------------------------------------------------------------------------%
% Section
%-------------------------------------------------------------------------------------------------%
%-------------------------------------------------------------------------------------------------%
\section{Derivation of the expression of the deterministic dynamic bifurcation point}\label{app:2}

In this appendix we give the details of the résolution of Eq.~\eqref{eq:AA0}. First we state $y+\hat{\gamma}^{\text{st}}=X^2$ and then it can be shown that, using Eq.~\eqref{eq:defA}, Eq.~\eqref{eq:AA0} takes the following form
\begin{multline}
\left(X-X_0\right)
\Big(\zeta  F_1 X^2+
(\zeta  F_1X_0-\alpha _1\omega_1)X\\
-\zeta  F_1+\zeta  F_1 X_0^2-\alpha _1 X_0 \omega _1\Big)=0
\label{eq:AA0b}
\end{multline}
where $X_0^2=y_0+\hat{\gamma}^{\text{st}}$. Obviously $X=X_0$ and therefore $y=y_0$ is a solution of \eqref{eq:AA0b}. The second term of the product in the left-hand side of \eqref{eq:AA0b} is a second order polynomial equations whose roots $X_1$ and $X_2$ are
\begin{align}
X_{1}&=\frac{1}{2 \zeta  F_1}\Bigg(\alpha _1 \omega _1-\zeta  F_1 X_0\nonumber\\
&-\sqrt{\alpha _1^2 \omega _1^2+2 \alpha _1 \zeta  F_1 X_0 \omega _1+\zeta ^2 F_1^2 \left(4-3 X_0^2\right)}\Bigg)
\end{align}
and
\begin{align}
X_2&=\frac{1}{2 \zeta  F_1}\Bigg(\alpha _1 \omega _1-\zeta  F_1 X_0\nonumber\\
&+\sqrt{\alpha _1^2 \omega _1^2+2 \alpha _1 \zeta  F_1 X_0 \omega _1+\zeta ^2 F_1^2 \left(4-3 X_0^2\right)}\Bigg).
\label{eq:X2}
\end{align}

The initial value $y_0$ is always chosen to be larger than $-\hat{\gamma}^{\text{st}}$ (because the mouth pressure $\gamma$ must be larger than zero). Therefore, $X$ must be larger than zero and, for a realistic set of parameters, only $X_2$ is positive. Consequently, the expression of the deterministic dynamic bifurcation point is given by
\begin{equation}
\hat{y}^{\text{dyn}}_{\text{det}}=X_2^2-\hat{\gamma}^{\text{st}}.
\label{eq:dyndetbifpt10}
\end{equation}

%-------------------------------------------------------------------------------------------------%
%-------------------------------------------------------------------------------------------------%
% Section
%-------------------------------------------------------------------------------------------------%
%-------------------------------------------------------------------------------------------------%
\section{Derivation of the expression of the stochastic dynamic bifurcation point}\label{app:stochDyn}

Using Eq.~\eqref{eq:defA}, Eq.~\eqref{eq:eqDynPtStoch2} takes the following explicit form
\begin{equation}
K=\frac{\zeta  F_1\sqrt{\hat{\gamma}^{\text{st}}  +y}}{2 \omega _1}\left(y-1+\hat{\gamma}^{\text{st}}\right)-\frac{\alpha _1 y}{2}.
\label{eq:eqDynPtStoch2b} 
\end{equation}

To obtain a solvable cubic form (i.e. without square root), Eq.~\eqref{eq:eqDynPtStoch2b} is transformed into 
\begin{equation}
\left(K+\frac{\alpha _1 y}{2}\right)^2=
\frac{\zeta^2  F_1^2\left(\hat{\gamma}^{\text{st}}  +y\right)}{4 \omega _1^2}
\left(y-1+\hat{\gamma}^{\text{st}}\right)^2
\label{eq:eqDynPtStoch2c} 
\end{equation}
which yields Eq.~ \eqref{eq:cubic}.

In the remaining of this appendix the Cardano's method (see e.g.~\cite{Spiegel2012}) is used to solve the latter, i.e.
$
a_1y^3+a_2y^2+a_3y+a_4=0.
$

First, the following parameters are introduced
$$
p=-\frac{a_2^2}{3a_1^2}+\frac{c_3}{a_1}\quad\text{and}\quad q=\frac{a_2}{27a_1}\left(\frac{2a_2^2}{a_1^2}-\frac{9a_3}{a_1}\right)+\frac{a_4}{a_1}.
$$
The discriminant $\Delta$ is defined as
$
\Delta=-\left(4 p^3+27 q^2\right).
$
Then:
\begin{enumerate}
\item If $\Delta<0$, one root is real and two are complex conjugate.
\item If $\Delta=0$, all roots are real and at least two are equal.
\item If $\Delta>0$, all roots are real and unequal.
\end{enumerate}

\begin{figure}[t!]
	\centering
	\includegraphics[width=1\columnwidth]{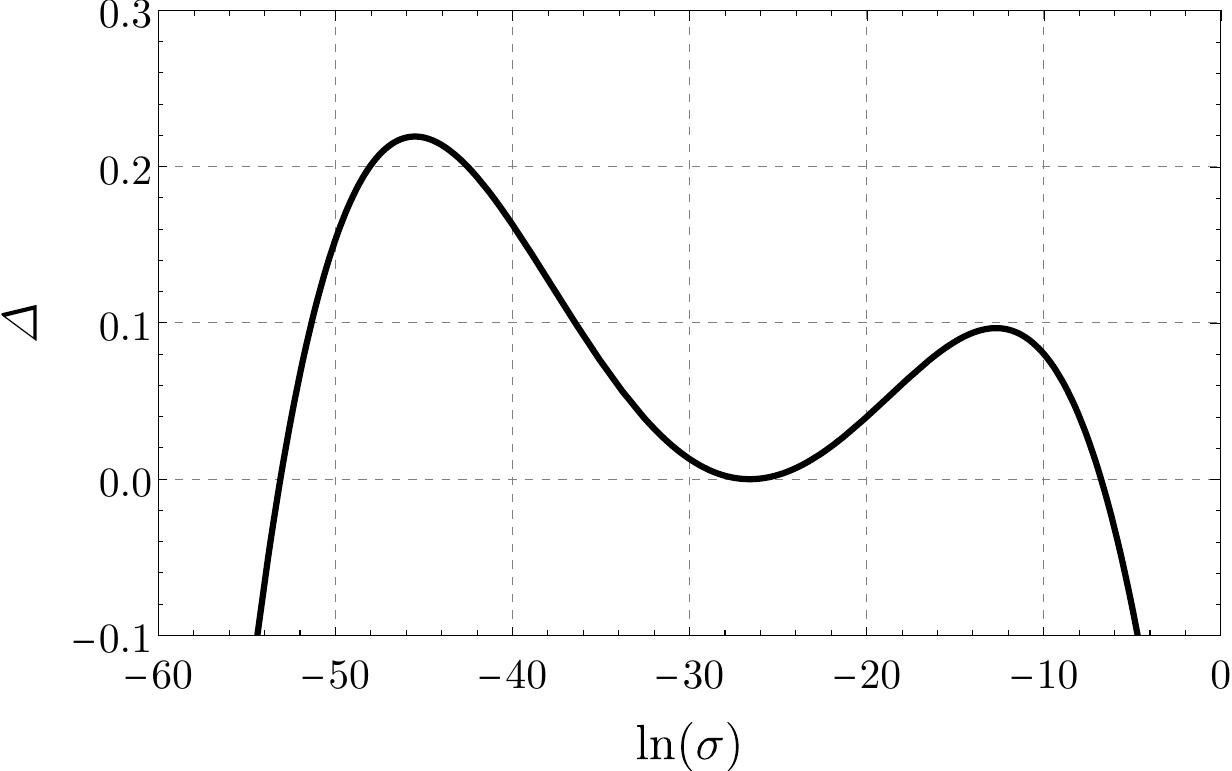}
	\caption{Discriminant $\Delta$ of \eqref{eq:cubic} as a function the natural logarithm of the noise level $\ln \sigma$. The set of parameters \eqref{eq:param1} is used.}
	\label{fig:pDelta}
	%param = {\[Alpha]1 -> 0.02`100, \[Omega]1 -> 1000`100,   F1 -> 1200`100, \[Epsilon] -> 0.002`100, \[Zeta] -> 0.2`100};
	%Nrealizations = 10
\end{figure}

A typical example of the discriminant $\Delta$, plotted as a function of the noise level $\sigma$, is depicted in Fig.~\ref{fig:pDelta} for a typical set of parameters. It can be shown that $\Delta$ can be expressed as a fourth order polynomial equation with respect to $\ln\sigma$ which has two distinct roots
\begin{align}
\ln\left(\sigma_1\right)&=\frac{2 \alpha _1^3 \omega _1^3-27 \alpha _1 \hat{\gamma}^{\text{st}}   \zeta ^2 F_1^2 \omega _1}{54 \zeta ^2 F_1^2 \omega _1 \epsilon }\nonumber\\
&-\frac{2 \left(\alpha _1^2 \omega _1^2+3 \zeta ^2 F_1^2\right)^{3/2}-9 \alpha _1 \zeta ^2 F_1^2 \omega _1}{54 \zeta ^2 F_1^2 \omega _1 \epsilon }\nonumber\\
&-\frac{1}{2} \ln \left(\frac{2 \sqrt{2 \pi } \hat{\gamma}^{\text{st}^{3/4}} \omega _1 e^{-\frac{(\hat{\gamma}^{\text{st}}  -1) \sqrt{\hat{\gamma}^{\text{st}}  } \zeta  F_1}{\omega _1 \epsilon }}}{\sqrt{(3 \hat{\gamma}^{\text{st}}  +1) \zeta  F_1 \omega _1 \epsilon }}\right)\nonumber\\
&+\ln \left(x_{y_0}\right)
\label{eq:sigma1}
\end{align}
and
\begin{align}
\ln\left(\sigma_2\right)&=\frac{9 \alpha _1 (1-3 \hat{\gamma}^{\text{st}}  ) +\frac{2 \left(\alpha _1^3 \omega _1^3+\left(\alpha _1^2 \omega _1^2+3 \zeta ^2 F_1^2\right){}^{3/2}\right)}{\zeta ^2 F_1^2 \omega _1}}{54 \epsilon}\nonumber\\
&-\frac{1}{2} \ln \left(\frac{2 \sqrt{2 \pi } \hat{\gamma}^{\text{st}^{3/4}} \omega _1 e^{-\frac{(\hat{\gamma}^{\text{st}}  -1) \sqrt{\hat{\gamma}^{\text{st}}  } \zeta  F_1}{\omega _1 \epsilon }}}{\sqrt{(3 \hat{\gamma}^{\text{st}}  +1) \zeta  F_1 \omega _1 \epsilon }}\right)\nonumber\\
&+\ln \left(x_{y_0}\right)
\label{eq:sigma2}
\end{align}
and a double root
\begin{align}
\ln\left(\sigma_3\right)&=\frac{\alpha _1 (\hat{\gamma}^{\text{st}} -1 )}{2 \epsilon }\nonumber\\
&-\frac{\log \left(\frac{2 \sqrt{2 \pi } \hat{\gamma}^{\text{st}^{3/4} }\omega _1 e^{-\frac{(\hat{\gamma}^{\text{st}}  -1) \sqrt{\hat{\gamma}^{\text{st}}  } \zeta  F_1}{\omega _1 \epsilon }}}{\sqrt{(3 \hat{\gamma}^{\text{st}}  +1) \zeta  F_1 \omega _1 \epsilon }}\right)}{2}\nonumber\\
&+ \ln \left(x_{y_0}\right).
\label{eq:sigma3}
\end{align}
In the example shown in Fig.~\ref{fig:pDelta} one has: $\ln\left(\sigma_1\right)=-53.1$, $\ln\left(\sigma_2\right)=-6.8$ and $\ln\left(\sigma_3\right)=-26.6$.

\begin{figure}[t!]
	\centering
	\includegraphics[width=1\columnwidth]{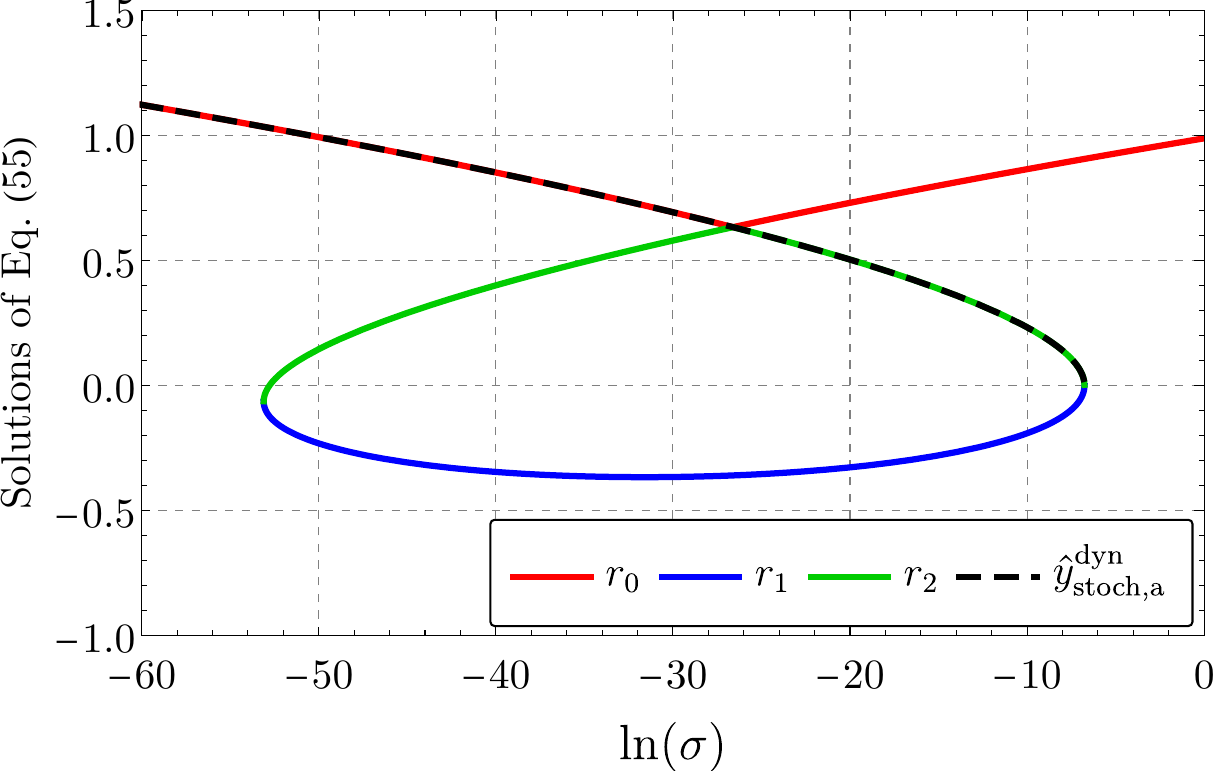}
	\caption{The roots $r_k$ ($k=0,1,2$) of Eq.~\eqref{eq:cubic} and the the stochastic dynamic bifurcation $\hat{y}^{\text{dyn}}_{\text{stoch,a}}$ as functions of the natural logarithm of the noise level $\sigma$. The parameters \eqref{eq:param1} are used and $x_{y_0}=0.01$.
	}
	\label{fig:pSoluCard}
\end{figure}

If $\sigma_1<\sigma<\sigma_2$ we have $\Delta>0$ except at $\sigma=\sigma_3$ for which $\Delta=0$. In general, when $\Delta>0$ the three real roots are written using trigonometric functions as follows
\begin{multline}
r_k = 2 \sqrt{\frac{-p}3} \cos{\left(\frac13\arccos{\left(\frac{3q}{2p}\sqrt{\frac3{-p}}\right)}+ \frac{2k\pi}3\right)}
-\frac{a_2}{3a_1}\\
\text{with}\quad k=0,1,2.
\label{eq:cardansol1}
\end{multline}
Fig.~\ref{fig:pSoluCard} shows the roots $r_k$ ($k=0,1,2$) as functions of the noise level $\sigma$ using again the parameters \eqref{eq:param1} and $x_{y_0}=0.01$. The stochastic dynamic bifurcation point, denoted $\hat{y}^{\text{dyn}}_{\text{stoch,a}}$, is equal to $r_0$ if $\sigma<\sigma_3$ and to $r_2$ if $\sigma_3<\sigma<\sigma_2$. This choice is justified by means of a numerical resolution which shows that this is the unique positive solution of Eq.~\eqref{eq:eqDynPtStoch2b}.

%-------------------------------------------------------------------------------------------------%
%-------------------------------------------------------------------------------------------------%
% Section
%-------------------------------------------------------------------------------------------------%
%-------------------------------------------------------------------------------------------------%
\section{Static bifurcation diagram}\label{app:bifdiag}

The static bifurcation diagram is the result of the bifurcation analysis of a deterministic dynamical system with constant bifurcation parameter, here Eq.~\eqref{eq:VdP1Mode1} with a constant value of $y$. It plots, as a function of the considered bifurcation parameter (here $y$), the possible steady-state regimes (fixed points and periodic motions) indicating their stability.

We use here the averaging procedure to obtain the approximated analytical bifurcation diagram of the one-mode model. For that Eq.~\eqref{eq:VdP1Moy3Stocha} is considered without noise et with a constant bifurcation parameter $y$, i.e.
\begin{equation}
\frac{dx_t}{dt}=F(x_t,y),
\label{eq:bifdiag1}
\end{equation}
where the function $F(x_t,y_t)$ is given by Eq.~\eqref{eqF1}.

The fixed points $x^e$ of \eqref{eq:bifdiag1} are obtained by solving 
$
F(x,y)=0.
$
We obtain three solutions: the trivial solution $x^e_1=0$ and two non trivial solutions, one is negative and one is positive. Only the the positive non trivial solution is retained and denoted $x^e_2$, its expression is
\begin{equation}
x^e_2=4\sqrt{\frac{\frac{2 \zeta  F_1 (3 \hat{\gamma}^{\text{st}} +3 y-1)}{3 \omega _1 \sqrt{\hat{\gamma}^{\text{st}} +y}}-\frac{4 \alpha _1}{3}}{\frac{\zeta  F_1 (\hat{\gamma}^{\text{st}} +y+1)}{\omega _1 (\hat{\gamma}^{\text{st}} +y)^{5/2}}}}
\label{eq:SBD}
\end{equation}
where the expression of $\hat{y}^{\text{st}}$ is given by Eq.~\eqref{eq:statbifpt}. In the lossless case with $\alpha_1=0$ and $\hat{\gamma}^{\text{st}}=\frac{1}{3}$, Eq.~\eqref{eq:SBD} reduces to $x^e_2=4 \sqrt{2} (3 y+1) \sqrt{\frac{y}{9 y+12}}$.

The trivial solution corresponds to zero equilibrium position of \eqref{eq:VdP1Mode1} whereas the non trivial solution characterizes its periodic steady-state regimes. 

As we know, the trivial fixed point $x^e_1$ is stable if $y<0$ and unstable if $y>0$. One can be shown that the non trivial fixed point $x^e_2$ exists only for $y>0$ and is stable. %The resulting bifurcation diagram is depicted in Fig.~\ref{fig:DiagBifStat}.

%-------------------------------------------------------------------------------------------------%
% Bibliographie
%-------------------------------------------------------------------------------------------------%
\small
%\bibliographystyle{spmpsci}  
%\bibliography{biblio_PIM}

\end{document}